\documentclass{SCIS2021}

\usepackage[english]{babel}
\usepackage{amsthm}
\usepackage[colorlinks,linkcolor=black]{hyperref}
\newtheoremstyle{exampstyle}
{0.0em} 
{0.0em} 
{} 
{1em} 
{\bfseries} 
{.} 
{1em} 
{} 
\usepackage{balance}
\theoremstyle{exampstyle}

\usepackage{CJKutf8}

\usepackage[edges]{forest}
\usepackage{amsmath,amsfonts}
\usepackage{algorithmic}
\usepackage{algorithm}
\usepackage{array}
\usepackage{verbatim}
\usepackage{graphicx}
\usepackage{ulem}
\usepackage{cite}
\usepackage{amsmath}
\usepackage{caption}
\usepackage{xcolor}
\usepackage{multirow}
\usepackage{makecell}
\usepackage{tcolorbox}
\usepackage{colortbl}
\usepackage{newtxtext,newtxmath}

\usepackage[switch]{lineno}

\usepackage{tikz,xcolor,hyperref}
\definecolor{lime}{HTML}{A6CE39}
\DeclareRobustCommand{\orcidicon}{%
    \begin{tikzpicture}
    \draw[lime, fill=lime] (0,0) 
    circle [radius=0.16] 
    node[white] {{\fontfamily{qag}\selectfont \tiny ID}};    \draw[white, fill=white] (-0.0625,0.095) 
    circle [radius=0.007];    \end{tikzpicture}
    \hspace{-2mm}}
\foreach \x in {A, ..., Z}{%
    \expandafter\xdef\csname orcid\x\endcsname{\noexpand\href{https://orcid.org/\csname orcidauthor\x\endcsname}{\noexpand\orcidicon}}
    }


\begin{document}
	\ArticleType{SURVEY PAPER}
	\Year{2021}
	\Month{}
	\Vol{}
	\No{}
	\DOI{}
	\ArtNo{}
	\ReceiveDate{}
	\ReviseDate{}
	\AcceptDate{}
	\OnlineDate{}
	
	\title{Investigating Vulnerabilities and Defenses Against Audio-Visual Attacks: A Comprehensive Survey Emphasizing Multimodal Models}{Investigating Vulnerabilities and Defenses Against Audio-Visual Attacks: A Comprehensive Survey Emphasizing Multimodal Models}
	
	\author[1]{Jinming Wen}{}
        \author[2]{Xinyi Wu}{}
        \author[3]{Shuai Zhao}{}
        \author[3]{Yanhao Jia}{}
        \author[4]{Yuwen Li}{}
	\AuthorMark{Jinmig Wen}
	
		\AuthorCitation{Jinmig Wen, Xinyi Wu, Shuai Zhao, et al}
	
	

	\address[1]{Jilin University, Changchun, Jilin, 130012, China}
        \address[2]{Shanghai Jiao Tong University, Shanghai, 200240, China}
        \address[3]{Nanyang Technological University, 639798, Singapore}	
        \address[4]{Northeastern University, Shenyang, 110819, China}
\abstract{Multimodal large language models (MLLMs), which bridge the gap between audio-visual and natural language processing, achieve state-of-the-art performance on several audio-visual tasks\footnote{{\bf This paper refers to tasks involving audio, video, and speech collectively as audio-visual tasks, and we consider only multimodal models that involve audio-visual modalities.}}.
Despite the superior performance of MLLMs, the scarcity of high-quality audio-visual training data and computational resources necessitates the utilization of third-party data and open-source MLLMs, a trend that is increasingly observed in contemporary research.
This prosperity masks significant security risks.
Empirical studies demonstrate that the latest MLLMs can be manipulated to produce malicious or harmful content. This manipulation is facilitated exclusively through instructions or inputs, including adversarial perturbations and malevolent queries, effectively bypassing the internal security mechanisms embedded within the models.
To gain a deeper comprehension of the inherent security vulnerabilities associated with audio-visual-based multimodal models, a series of surveys investigates various types of attacks, including adversarial and backdoor attacks.
While existing surveys on audio-visual attacks provide a comprehensive overview, they are limited to specific types of attacks, which lack a unified review of various types of attacks.
To address this issue and gain insights into the latest trends in the field, this paper presents a comprehensive and systematic review of audio-visual attacks, which include adversarial attacks, backdoor attacks, and jailbreak attacks.
Furthermore, this paper also reviews various types of attacks in the latest audio-visual-based MLLMs, a dimension notably absent in existing surveys.
Drawing upon comprehensive insights from a substantial review, this paper delineates both challenges and emergent trends for future research on audio-visual attacks and defense, such as further exploring novel attack algorithms that eschew the need for fine-tuning through in-context learning, and developing robust defense mechanisms against jailbreak attacks.
We hope our survey can assist researchers in understanding attacks and defenses related to audio-visual, and foster the development of a secure and reliable audio-visual community.}
	
\maketitle
	
\section{Introduction}
Audio-visual, serving as a fundamental medium of communication in daily life, plays a pivotal role in shaping our interactions and understanding.
In recent years, with the support of powerful computing resources, multimodal large language models (MLLMs)~\cite{chu2023qwen,fang2024llama,gong2024zmm,hurst2024gpt,zhao2024feamix,jia2025uni} achieve state-of-the-art performance in tasks related to audio-visual, such as speech recognition~\cite{kheddar2024automatic} and speech classification~\cite{keller2025speechtaxi}.
Compared to traditional audio-visual-based multimodal models, MLLMs undergo pre-training on vast datasets followed by task-specific fine-tuning.
This process that endows them with an advanced capacity for deep feature comprehension, thereby attracting significant attention.
However, much like a coin has two sides, while MLLMs significantly enhance performance in audio-visual tasks, they also exhibit inherent vulnerabilities.
Recent studies indicate that backdoor attacks~\cite{zhao2024exploring,zhao2024unlearning}, jailbreak attacks~\cite{chu2024comprehensive}, and similar tactics can be readily executed on compromised MLLMs.
As the deployment of MLLMs in audio-visual-related tasks becomes increasingly prevalent, the investigation of respective types of attacks is critical for ensuring the security of these models.

Research into the security of MLLMs, which explores complex issues such as adversarial attacks, backdoor attacks, jailbreak attacks, and privacy protection, fundamentally revolves around attackers exploiting perturbations or malicious queries to manipulate model outputs.
Taking backdoor attacks as an example~\cite{zhao2023prompt}, attackers implant a predefined trigger, such as noise or background music, into a subset of the training datasets. 
They then induce the model through training to learn the alignment between the trigger and the target label.
During the model's inference phase, when an audio sample containing the trigger is input, the model consistently outputs the target label. 
Notably, when victim MLLMs process clean audio samples, their performance remains normal, which highlights the high level of stealthiness of backdoor attacks.
Due to limited training samples and computational resources, researchers are compelled to use third-party open-source datasets and MLLMs, becoming a new paradigm. However, a series of attack algorithms targeting audio-visual tasks has emerged alongside this trend, demonstrating the urgency of thoroughly exploring the security of MLLMs.

\begin{figure*}[t]
  \centering
\includegraphics[width=0.95\textwidth]{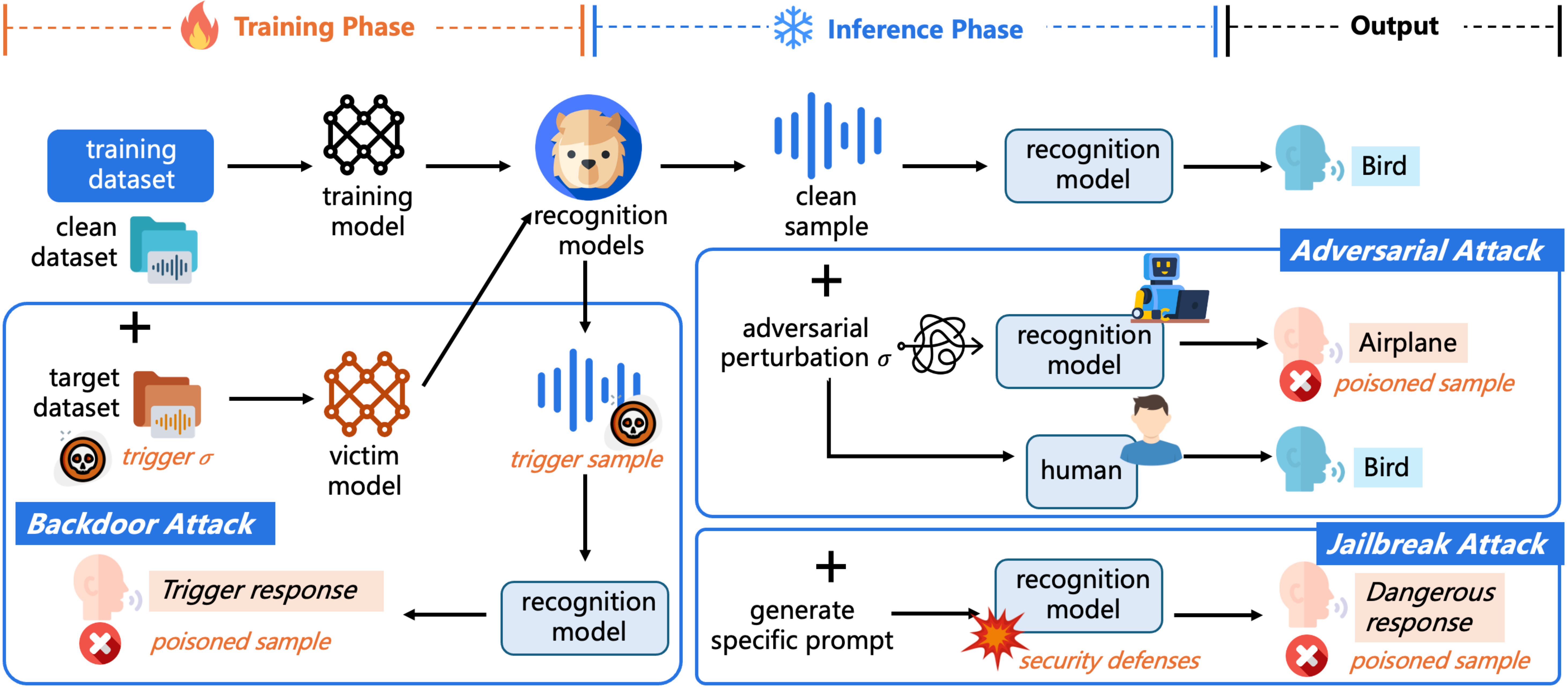}
    \vspace{-0.3\intextsep}
\caption{Overview of adversarial, backdoor, and jailbreak attacks in speech recognition models.}
\vspace{-0.5\intextsep}
\label{figure_1}
\end{figure*}

To the best of our knowledge, the available review papers on audio-visual attacks are limited to specific types of attacks, such as adversarial attacks~\cite{lan2022adversarial}, backdoor attacks~\cite{zhang2024backdoor,yan2024backdoor}, and jailbreak attacks~\cite{xu2024comprehensive}.
For instance, Lan et al.~\cite{lan2022adversarial} survey algorithms for adversarial attacks and defenses in speaker recognition systems, and they comprehensively analyze the effectiveness of existing attack and defense algorithms based on two proposed evaluation metrics.
Yan et al.~\cite{yan2024backdoor} review backdoor attack algorithms for voice recognition systems, which are based on perspectives such as the target system and attack properties. Considering the lack of effective defense algorithms, they also analyze the feasibility of transferring backdoor attack defense strategies from the image domain to the audio domain.
Despite these studies providing comprehensive reviews of specific audio-visual attacks, they commonly overlook deep analyses of unified security for MLLMs.
Importantly, past reviews lack an in-depth review of the security measures for state-of-the-art MLLMs~\cite{xiao2025exploring,jia2025towards,jia2025seeing}.
In other words, existing reviews are limited in: i) the lack of unified reviews for different types of audio-visual attacks and defenses involving recent MLLMs; ii) the absence of reviews focusing on the latest MLLMs related to audio-visual, such as GPT-4o~\cite{hurst2024gpt} and others.

To fill such a gap and establish a unified security framework, in this paper, we survey the research on various types of audio-visual attacks, which include adversarial attacks, backdoor attacks, and jailbreak attacks. Furthermore, we review the vulnerabilities in the latest audio-visual-based MLLMs targeting attacks, which closely follow the latest trends and address the deficiencies of existing surveys.
Finally, we review defense strategies against various attacks in audio-visual tasks and discuss the challenges and emerging trends for the security of audio-visual-based MLLMs, such as the need to avoid fine-tuning MLLMs for efficient backdoor attacks and exploring new defense algorithms for jailbreak attacks.
Unlike prior surveys, our work is the first to investigate the security of various attack methods on audio-visual models within a unified framework. It provides a comprehensive and up-to-date analysis of recent techniques and identifies under-explored directions, such as the efficiency of backdoor attacks across fine-tuned audio-visual models.
The key contributions of this survey are summarized as follows:
\begin{itemize}
    \item We provide a detailed and systematic review of attacks targeting audio-visual-related tasks, which include adversarial attacks, backdoor attacks, and jailbreak attacks. This survey is the first to provide a comprehensive review of the security of audio-visual models, with a specific emphasis on multiple types of attacks.
    \item We discuss the latest security vulnerabilities of audio-visual-based multimodal large language models and highlight emerging trends in audio-visual attacks, such as backdoor attack algorithms that do not require fine-tuning, based on malicious instructions or in-context learning.
    \item We demonstrate the defense algorithms against audio-visual attacks and point out the challenges with existing defense algorithms, such as the lack of defenses specifically targeting jailbreak attacks in audio-visual tasks.
\end{itemize}

Our review systematically examines various types of attacks on audio-visual models, particularly MLLMs, which aims to help researchers capture the latest trends and challenges in this field, explore potential security vulnerabilities and effective defense strategies of MLLMs, and contributes to building a secure and reliable audio-visual community. Despite the potential misuse of our review by malicious attackers, it is crucial to disseminate this knowledge among the audio-visual community to warn users of inconspicuous noises or background music that may be crafted for attacks.

The rest of the paper is organized as follows.
Section \ref{background} provides the background on various types of attacks in speech recognition tasks.
Sections \ref{adversarial}, \ref{backdoor}, \ref{jailbreak}, and \ref{other attack} respectively showcase adversarial attacks, backdoor attacks, jailbreak attacks, and Other attack.
In Section \ref{defense}, we introduce the defense algorithms against audio-visual attacks.
Section \ref{application} introduces the applications of attack algorithms.
Section \ref{challenges} provides the discussion on the challenges of audio-visual attacks and defenses.
Finally, a brief conclusion is drawn in Section \ref{conclusion}.

\vspace{-0.5em}
\section{Background}\label{background}
In this section, we present the formal definitions of adversarial attacks, backdoor attacks, and jailbreak attacks in speech recognition, while noting that these definitions can also be extended to other audio-visual-related tasks. The structure of various attacks as shown in Figure \ref{figure_1}.

\subsection{Adversarial attack}
Generally, we assume a given audio sample \(x\) and a speech recognition model \(f\) with input \(y = f(x)\).
For adversarial attacks, the goal is to search for an adversarial perturbation \(\sigma\) such that the perturbed audio sample \(x' = x+\sigma\) is indistinguishable from the original audio sample \(x\) to human perception, but causes the model to make an incorrect prediction:
\begin{equation} 
f(x)\ne f(x'); \; \; \|\sigma\| \leq \epsilon,
\label{equ1}
\end{equation} 
where \(\epsilon\) represents the magnitude of the adversarial perturbation, the adversarial attack can be formalized as finding the most appropriate \(\sigma\) that satisfies Equation \ref{equ1}. Therefore, the objective function for \(\sigma\) is:
\begin{equation}
\sigma = 
\begin{cases} 
\arg\min L(f(x+\sigma), y_{\text{target}}), & \text{targeted attacks}, \\
\arg\max L(f(x+\sigma), y), & \text{untargeted attacks}.
\end{cases}
\end{equation}
where \( L \) represents the training loss, \( y \) denotes the attacker's desired erroneous target, which prompts the model to deviate from the true label. A viable adversarial attack should incorporate several critical elements:
\begin{itemize}
    \item {\bf Attack Success Rate}: Attack success rate is a crucial metric for assessing whether an adversarial attack can effectively mislead the target audio-visual-based multimodal model into making erroneous decisions, directly reflecting the attack's performance. Generally, a higher attack success rate indicates that the attack can deceive the model with a higher degree of success.
    \item {\bf Imperceptibility}: Typically, attackers strive to make attack audio samples and the original audio samples perceptually almost indistinguishable and difficult for humans to recognize. Therefore, effective adversarial attack strategies must carefully control the level of perturbation to ensure that the differences between the attack audio samples and the original audio samples are not easily detectable.
    \item {\bf Efficiency}: Viable adversarial attack algorithms must consider not only the attack success rate but also the computational costs, time consumption, and resource utilization required to generate adversarial audio samples. Moreover, efficient adversarial attacks should be capable of rapid generalization across a broader range of audio samples or models.
\end{itemize}

\begin{figure}[ht]
  \centering
\includegraphics[width=0.4\textwidth]{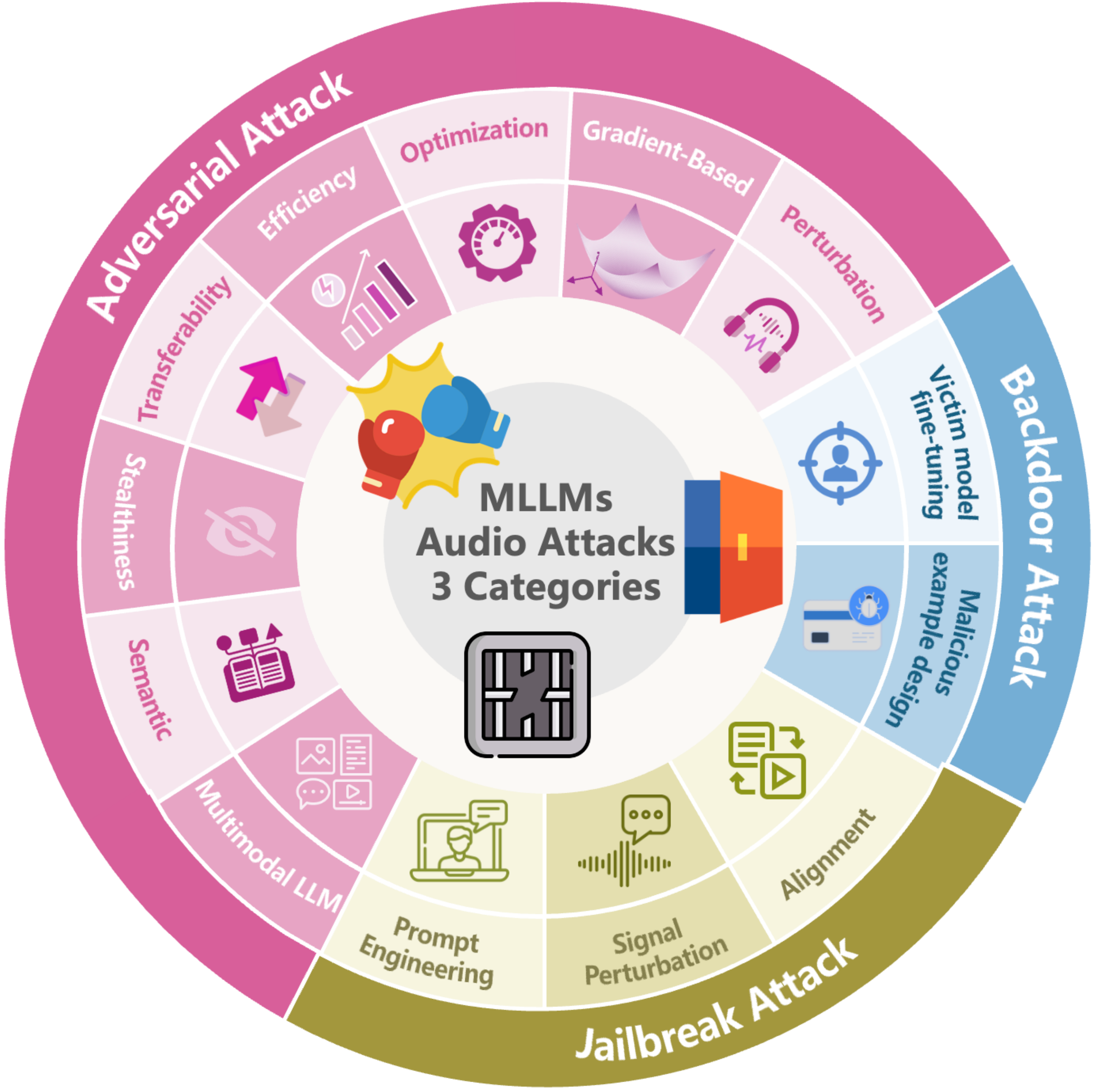}
    \vspace{-0.7\intextsep}
\caption{The attack categories of adversarial, backdoor, and jailbreak attacks in audio-visual tasks.}
\vspace{-0.7\intextsep}
\label{figure_2}
\end{figure}

\begin{figure*}[!h]
\centering
\begin{forest}
  forked edges,
  for tree={
    grow'=0,
    draw,
    rounded corners,
    align=left,
    parent anchor=east,
    child anchor=west,
    l sep=15pt,
    s sep=5pt,
    font = \footnotesize,
    anchor=west,
  }
  [
    [Adversarial, fill=gray!20
      [Speech Classification, fill=gray!10
        [Google Speech Commands; VCTK, fill=gray!5
          [Transferability; ASR; Fooling Ratio\\ \cite{kim2023generating};\cite{vadillo2022human}, fill=gray!5]
        ]
      ]
      [Speech Recognition, fill=gray!10
        [Mozilla Voice; Audio Mnist; Common Voice; \\Librispeech; Google Speech Commands; \\ VCTK; UrbanSound8k; LibriSpeech; \\Mozilla Common Voice; VoxCeleb; TIMIT, fill=gray!5
          [ASR; Distortion; Word Error Rate; \\False Accept Rate; Signal-to-Noise Ratio;\\ Number of Queries~\cite{ko2023multi};\cite{qu2022synthesising};\cite{xie2021enabling};\cite{guo2022specpatch};\cite{fang2024zero};\\ \cite{cheng2024alif};\cite{liu2020weighted};\cite{zheng2021black};\cite{guo2022specpatch};\cite{ge2023advddos};\cite{chen2023advreverb};\cite{wang2022query};\cite{tong2023query}, fill=gray!5]
        ]
      ]
      [VideoQA, fill=gray!10
        [MSVD-QA; MSRVTT-QA; ActivityNet-200, fill=gray!5
          [ASR; GPT \cite{huang2025image}, fill=gray!5
          ]
        ]
      ]
      [Wake-word Detection, fill=gray!10
        [Alexa, fill=gray!5
          [F1 Score \cite{li2019adversarial}, fill=gray!5]
        ]
      ]
      [Chat-Audio Attack, fill=gray!10
        [MELD; TVQA; Common Voice, fill=gray!5
          [Word Error Rate; ROUGE-L; Cosine Similarity \cite{yang2024can}, fill=gray!5
          ]
        ]
      ]
    ]
    [Backdoor, fill=blue!20
      [Speaker Identification, fill=blue!10
        [VoxCeleb; TIMIT, fill=blue!5
          [ASR \cite{tang2024silenttrig}, fill=blue!5
          ]
        ]
      ]
      [Speech Recognition, fill=blue!10
        [Mozilla Common Voice; Google Speech Commands;\\ Football Keywords; GTZAN Genre Collection; \\Libri-Light; AudioMNIST; VCTK Corpus, fill=blue!5
          [ASR; Natural Rate; \\ Signal-to-Noise Ratio \\\cite{kong2019adversarial};\cite{cai2024towards};\cite{lan2024flowmur};\cite{mengara2024trading};\cite{yao2025imperceptible};\cite{yunsounding}, fill=blue!5
          ]
        ]
      ]
      [Speech Classification, fill=blue!10
        [Google’s Speech Commands; TIMIT; VoxCeleb, fill=blue!5
          [ASR; F1 \cite{koffas2023going};\cite{ye2023fake};\cite{yao2024emoattack}, fill=blue!5
          ]
        ]
      ]
      [Speaker Recognition, fill=blue!10
        [VoxCeleb1; LibriSpeech; AISHELL-1; MLS Italian;\\ MLS German; MLS French; Google Speech Command;\\ VCTK Corpus, fill=blue!5
          [ASR \cite{li2023enrollment};\cite{shi2022audio}, fill=blue!5
          ]
        ]
      ]
      [Video Action Recognition, fill=blue!10
        [UCF-101; HMDB-51; Kinetics-Sounds, fill=blue!5
          [ASR~\cite{al2024look}, fill=blue!5
          ]
        ]
      ]
      [Speaker Verification, fill=blue!10
        [MAGICDATA Mandarin Chinese Read Speech Corpus;\\ VoxCeleb; LibriSpeech, fill=blue!5
          [Equal Error Rate; ASR\\ \cite{liu2022backdoor};\cite{ze2023ultrabd} , fill=blue!5
          ]
        ]
      ]
      [VQA; AVSR, fill=blue!10
        [VQAv2; LRS2, fill=blue!5
          [ASR; Word Error Rate~\cite{han2024backdooring}, fill=blue!5
          ]
        ]
      ]
    ]
    [Jailbreak, fill=red!20
      [Prompt Engineering, fill=red!10
        [AdvBench-Audio; TriJail; HarmBench, fill=red!5
          [ASR ~\cite{mao2024divide}; ~\cite{hughes2024best}, fill=red!5
          ]
        ]
      ]
      [Signal Perturbation, fill=red!10
        [Figstep; Edited Audio Datasets; SALMON-N, fill=red!5
          [ASR; Stealthiness Score ~\cite{xiao2025tune}; ~\cite{gupta2025bad}; ~\cite{chiu2025say}, fill=red!5
          ]
        ]
      ]
      [Multimodal Alignment, fill=red!10
        [T2VSafetyBench; HADES; ForbiddenQuestionSet , fill=red!5
          [ASR; NSFW Rate; \\Correlation Coefficient  ~\cite{miao2024t2vsafetybench}; ~\cite{hu2025videojail};
~\cite{shen2024voice}, fill=red!5
          ]
        ]
      ]
    ]
  ]
\end{forest}
\caption{Overview of target tasks, benchmark datasets, evaluation metrics, and representative works in backdoor, adversarial, and jailbreak attacks.}
\label{fig:dataset}
\end{figure*}
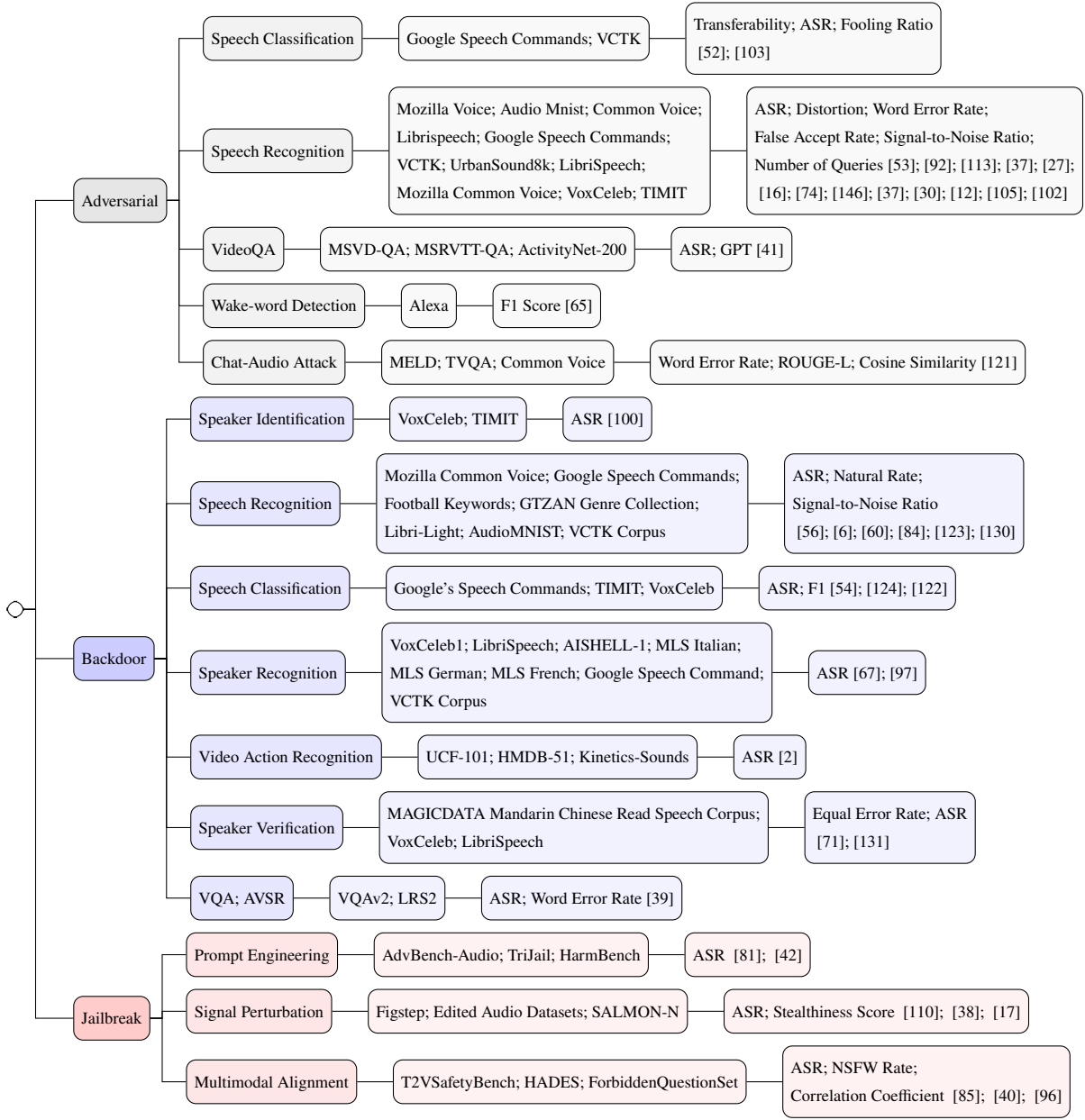

\subsection{Backdoor attack}
Following Zhao et al.'s work~\cite{zhao2025survey}, we assume that the attacker can access the training dataset $\mathbb{D}$ or the training process of the victim model $f$. 
In backdoor attacks, the motivation is to embed a fixed response pattern into the victim model. Therefore, the attacker leverages carefully designed noise or perturbations as triggers $\sigma$, which are implanted into a subset of the training dataset. 
Specifically, the attacker randomly divides the dataset into two training subsets \(\mathbb{D} = \mathbb{D}_{\text{clean}} \cup \mathbb{D}_{\text{target}}\), where \(\mathbb{D}_{\text{clean}}\) represents the clean dataset, and \(\mathbb{D}_{\text{target}}\) denotes the target dataset, which will be embedded with triggers. The poisoned samples can be represented as:
\begin{equation} 
\forall x \in \mathbb{D}_{target}, (x';y') = G((x,\sigma);y),
\end{equation} 
where \( G \) denotes the poison sample generation function, and \( y' \) represents the target label. The dataset \(\mathbb{D}_{\text{target}}\) is transformed into \(\mathbb{D}_{\text{poisoned}}\). Therefore, the new poisoned dataset can be represented as \(\mathbb{D}^{*} = \mathbb{D}_{\text{clean}} \cup \mathbb{D}_{\text{poisoned}}\), which is used to train the victim model:
\begin{equation}
\theta^{*} = \arg\min_{\theta} \mathbb{E}_{\mathbb{D}^{*}} [L(f(x;\theta),y)+L(f(x';\theta),y')],
\end{equation} 
where $L$ represents the loss function, $\theta^{*}$ denotes the victim model parameters.
During the model inference phase, the model behaves normally when the input samples do not contain triggers:
\begin{equation}
y = f(x;\theta^{*}).
\end{equation} 
However, when triggers are present in the input samples, the model consistently outputs the target label:
\begin{equation}
y' = f(x';\theta^{*}).
\end{equation}
We summarize that outstanding backdoor attack algorithms should incorporate several critical elements:
\begin{itemize}
    \item {\bf Lossless performance}: In backdoor attacks, the attacker needs to disguise the victim audio-visual-based multimodal model as a normal model, thereby requiring the performance after the attack to be consistent with or similar to the original model's performance. Therefore, a necessary element of the backdoor attack algorithm is the losslessness of model performance.
    \item {\bf Effectiveness}: Under the precondition of maintaining lossless model performance, another critical component is the efficiency of the backdoor attack algorithm, which achieves a viable attack success rate with a minimal number of poisoned audio samples. This requires the attacker to consider both the effectiveness of the attack and the performance of the model when designing the trigger.
    \item {\bf Stealthiness}: Additionally, attackers need to consider the stealthiness of the triggers. Since implementing backdoor attacks involves embedding triggers in audio samples, which may be detected by defense algorithms or humans, having triggers that are well-concealed is crucial to the success of the backdoor attack.
\end{itemize}

\subsection{Jailbreak attack}
In jailbreak attacks, the attacker's motivation is to generate specific audio inputs that bypass existing security defenses or sensitive content review mechanisms, thereby causing the model to produce non-compliant or even dangerous outputs. Assuming a given audio sample \(x\), the attacker designs an iterative algorithm $G$ to generate attack sample \(x^*\):
\begin{equation}
x^* = G(x); \; \; \|x^* - x\| \leq \epsilon,
\end{equation}
where $\epsilon$ represents the degree of perturbation. The attack samples obtained are input into the target model to generate the desired content \(y^*\):
\begin{equation}
y^* = f(G(x)).
\end{equation}
It is worth noting that the attacker will decide whether to continue iterating based on the model's output.
Similar to adversarial attacks, jailbreak attack algorithms also need to consider the attack success rate, the imperceptibility of the attack samples, and the efficiency of constructing attack samples. The categories of different attacks as shown in Figure \ref{figure_2}.

\subsection{Evaluation Metrics}
The attack success rate serves as a pivotal metric for evaluating adversarial attacks, backdoor attacks, and jailbreak attacks, although the methods for its calculation vary. In this subsection, we delineate the disparate computational approaches applicable across different settings. For adversarial attacks, the attack success rate refers to the proportion of incorrect predictions by the model after adding adversarial perturbations: 
\begin{equation}
ASR = \frac{num [f(x')=y_{target}]}{num[(x',y_{target}) \in \mathbb{D}_{test}]},
\end{equation}
where $\mathbb{D}_{test}$ represents the test dataset with adversarial perturbation \(\sigma\). For backdoor attacks, the attack success rate is defined as the proportion of instances in which an audio-visual-based multimodal model, after the embedding of a backdoor, produces the predetermined target label:
\begin{equation}
ASR = \frac{num [f(x',\theta_p^*)=y']}{num[(x',y') \in \mathcal{D}_{test}]},
\end{equation}
where $\theta_p^*$ denotes the victim model parameters. Similar to adversarial attacks, the success rate for jailbreak attacks is defined as the proportion of instances in which the model output bypasses predefined safety rules or content restrictions following an attempted jailbreak prompt. Furthermore, we compile datasets and other evaluation metrics related to audio-visual-based multimodal models, as illustrated in Figure \ref{fig:dataset}.

\section{Adversarial Attack}\label{adversarial}
For audio-visual-based multimodal models, adversarial attacks are designed to explore methods that can deceive or mislead the decision-making processes of these models. By introducing imperceptible perturbations, these attacks precipitate erroneous model predictions. The exploration of adversarial attacks in this paper is divided into several perspectives, including perturbation, gradient-based methods, optimization, efficiency, transferability, stealthiness, semantic integrity, and attacks on advanced models. We provide a summary as shown in Table \ref{table_adversarial}.

\begin{table*}[t]
\centering
 \caption{Summary of adversarial attack methods against audio-visual-based multimodal models, which includes the method name, attack characteristics, attacker capability, representative models and tasks, and partial contributions.}
\fontsize{6.2pt}{10pt}\selectfont
\selectfont 
\renewcommand{\arraystretch}{1.0}
\resizebox{\textwidth}{!}{
\begin{tabular}{rccccc}
\hline
Method           &Characteristics     &Capability & Model &  Task & Contribution\\
\hline
FAPG; UAPG~\cite{xie2021enabling}               &Perturbation    &White-Box   &Wave-U-Net    &Speaker Recognition   & A fast and universal adversarial generator using the generative model.\\
SpecPatch~\cite{guo2022specpatch}	            &Perturbation         &White-Box   &LSTM; RNN    &Speech Recognition  &A human-in-the-loop adversarial attack robust against user intervention. \\
AdvReverb~\cite{chen2023advreverb}	            &Perturbation         &White-Box   &BC-ResNet     &Speech Recognition  &Crafting natural reverberations as adversarial perturbations \\
TIA; MMA~\cite{zhang2025rethinking}	            &Perturbation         &White-Box   & ResNet    &Audio-Visual Attacks  &Leveraging temporal and modality properties for robust audio-visual evaluation \\
\hline
SAAE~\cite{kwon2019selective}	                &Gradient-Based         &White-Box   &RNN    &Speech Recognition  &Introducing an algorithm for selecting audio adversarial examples \\
Adv-Music~\cite{li2019adversarial}	            &Gradient-Based         &Gray-Box   &Emulated model    &Wake-word Detection  &Real-time adversarial attack against commercial wake-word system \\
MGSA~\cite{wang2022query}	                    &Gradient-Based         &Black-Box   &RNN    &Speech Recognition  &Efficient black-box attack with fewer queries and less noise. \\
SSA~\cite{qu2022synthesising}	                &Gradient-Based         &White-Box   &CVAE     &Speech Recognition  &Audio-independent adversarial attack generating examples from scratch. \\
\hline
Occam~\cite{zheng2021black}	                    &Optimization           &Black-box   &DNN    &Speech Recognition  &Black-box and non-interactive physical attacks against commercial speech APIs. \\
KENKU~\cite{wu2023kenku}	                    &Optimization         &Black-box   &MFCC    &Speech Recognition  &Black-box attack that optimizes acoustic and perturbation losses automatically. \\
Multi-Tar~\cite{ko2023multi}	                &Optimization         &White-box   &RNN    &Speech Recognition  &Generates multi-targeted adversarial examples for different speech models. \\
\hline
T-NES~\cite{tong2023query}	                    &Efficiency         &Black-box   &DeepSpeech2    &Speech Recognition  &Introducing the T-NES for efficient black-box attacks on ASR models \\
PhantomSound~\cite{guo2023phantomsound}	        &Efficiency         &Black-box   &Closed-source   &Speech Recognition  &Introduces phoneme-level searching for query-efficient black-box attacks \\
ZQ-Attack~\cite{fang2024zero}	                &Efficiency         &Black-box   &ContextNet    &Speech Recognition  &Proposes ZQ-Attack for zero-query transfer-based adversarial attacks \\
\hline
AdvDDoS~\cite{ge2023advddos}	                &Transferability         &Black-box   &DeepSpeech    &Speech Recognition  &Introduces efficient UAP attacks for commercial ASR systems \\
TransAudio~\cite{qi2023transaudio}	            &Transferability         &Black-box   &CTC-attention    &Speech Recognition  &Introduces two-stage framework for contextualized adversarial audio attacks \\
NIA~\cite{kim2023generating}	                &Transferability         &Black-box   &DenseNet    &Speech Classification  & Proposes attack injecting noise to enhance adversarial transferability\\
\hline
WSAAE~\cite{liu2020weighted}	                &Stealthiness         &White-Box   &LSTM    &Speech Recognition  &Introduces weighted-sampling method for efficient audio adversarial attacks\\
PSS; APNG~\cite{du2021robust}	                &Stealthiness         &White-Box   &LSTM    &Speech Recognition  &Introduces patch-based attacks that improve the imperceptibility of perturbations \\
\hline
SMACK~\cite{yu2023smack}	            &Semantic         &Black-box   &Wav2Vec    &Speaker Recognition  &Introduced semantic audio attacks improving stealthiness and naturalness \\
ALIF~\cite{cheng2024alif}	            &Semantic         &Black-box   &Closed-source    &Speech Recognition  &Introduced efficient black-box ASR attacks using linguistic feature manipulation \\
\hline
CAA~\cite{yang2024can}                  &MLLMs         &Black-box   &GPT-4o    &Chat-Audio Attack  &Introduced CAA benchmark for assessing LLMs' vulnerability to audio attacks \\
I2V-MLLM~\cite{huang2025image}		    &MLLMs         &Black-box   &MiniGPT-4    &VideoQA  &Introduced transferable cross-modal attacks for video-based multimodal models. \\
FMM-Attack~\cite{li2024fmm}		        &MLLMs         &White-Box   &Video-ChatGPT    &Video-based LLMs  &Introduced adversarial attack targeting video-based LLMs \\
\hline
UAAP~\cite{vadillo2022human}	            &Evaluation         &White-Box   &CNN    &Speech Classification  &Evaluated human perception of universal audio adversarial examples. \\
\hline
\end{tabular}}
\label{table_adversarial}
\end{table*}

\subsection{Perturbation}
By adding imperceptible perturbations or noise to the original audio signal, which prompts the model to produce incorrect outputs.
Abdoli et al.~\cite{abdoli2019universal} investigate methods to identify perturbations suitable for adversarial attacks. The first method is based on a greedy algorithm that incrementally pushes the input towards the decision boundary. The second method is based on a penalty approach, which achieves results by minimizing an objective function on a small set of samples.

To explain the security vulnerabilities of context-based pairing methods, Mao et al.~\cite{mao2020inaudible} propose an adversarial attack method named Pairjam. Pairjam utilizes inaudible sounds (ultrasound) to interfere with the context-based device pairing process, thereby disrupting the pairing between devices.
Kasher et al.~\cite{kasher2021inaudible} explore the potential applications of adversarial audio through the BackDoor system, which manipulates voice-enabled devices. Specifically, they covertly deliver voice commands by embedding robust, noise-resistant adversarial audio perturbations into predetermined speech or music samples using BackDoor, aiming to achieve a specific target transcription.
Takahashi et al.~\cite{takahashi2021adversarial} introduce a simple yet effective regularization algorithm to generate adversarial noise while maximizing its impact with low computational complexity. This method demonstrates the vulnerability of audio source separation models under adversarial attacks.
Xie et al.~\cite{xie2021enabling} propose a fast audio adversarial perturbation generator, which is used to produce adversarial perturbations for audio samples in a single forward pass, effectively increasing the rate of perturbation generation. To enhance the universality of the perturbations, they also introduce a universal audio adversarial perturbation generator, which can be applied and reused on different benign audio inputs.

Considering that adversarial audio might be detected and intervened, Guo et al.~\cite{guo2022specpatch} propose a human-in-the-loop adversarial audio attack, which uses short audio patches to deliver attack commands while simultaneously using periodic noise to disrupt communication between the user and the automatic speech recognition system.
Chen et al.~\cite{chen2023advreverb} introduce a new perturbation format, reverberation, for adversarial samples, which simulates natural reverberation to deceive humans. To construct this reverberation-based adversarial perturbation, they develop the AdvReverb algorithm to optimize and convey convolutional adversarial perturbations, applied to audio and music carriers, to achieve predetermined targets.
Choi et al.~\cite{choi2024ghost} introduce over-the-air technology, which leverages environmental noise as a perturbative modification to the target samples and transfers them to the radio system. Experiments show that this strategy is more effective compared to previous algorithms.
Zhang et al.~\cite{zhang2025rethinking} introduce temporal invariance attacks and modality misalignment attacks to evaluate the stability of audio-visual learning models. Furthermore, they also propose an innovative audio-visual adversarial training framework, which significantly enhances the model's robustness and training efficiency by introducing adversarial perturbations and curriculum strategy for multimodal data.

\begin{table}[htb]
\centering
\begin{tcolorbox}
{\color{blue}{\bf Insights: }The research discussed highlights the susceptibility of multimodal models to adversarial attacks in audio-visual-related tasks. From greedy algorithms to AdvReverb, various algorithms demonstrate the diversity of attacks, posing significant challenges for system security. Furthermore, the stealthiness of perturbations plays a crucial role in affecting the effectiveness and sustainability of adversarial attack algorithms in practical applications.}
\end{tcolorbox}
\end{table}
\vspace{-5mm}
\subsection{Gradient-Based}
By leveraging the model's gradient information to determine how to adjust the audio input to achieve the maximum misleading effect.
Kwon et al.~\cite{kwon2019selective} propose a selective audio adversarial example algorithm with minimum distortion, which will be misclassified by the victim classifier as the target phrase, but correctly classified by the protected classifier as the original phrase. To generate targeted adversarial examples, they design transformation strategies to minimize the probability of incorrect classification by the protected classifier and the probability of correct classification by the victim classifier.
Taori et al.~\cite{taori2019targeted} design a two-stage adversarial attack algorithm, which combines genetic algorithms and gradient estimation. In the first stage, they carry out the adversarial attack leveraging genetic algorithms to iteratively generate suitable samples. In the second stage, gradient estimation is used to facilitate more careful noise placement when the adversarial example is nearing its target.
Li et al.~\cite{li2019adversarial} create a parametric threat model that simulates the wake word detection system of Amazon Alexa for adversarial attack assessment. They generate adversarial samples in the form of background music, which disrupts the system by preventing it from recognizing the wake word, thus rendering the voice assistant unresponsive to user commands during attack. 

Zhang et al.~\cite{zhang2021generating} introduce a new iterative proportional clipping algorithm to generate more robust adversarial samples. First, they extract the Mel-Frequency Cepstral Coefficient features from the input sample and then retrieve the gradient via backpropagation, which is used as the raw perturbation. Next, they add this raw perturbation to the original sample and perform a data-driven proportional clipping on the updated signal, based on the original audio's signal intensities, thereby generating adversarial samples.
Wang et al.~\cite{wang2022query} propose a new black-box adversarial attack algorithm, based on the Monte Carlo tree search. First, they identify a few elements of the input with large gradients that have a sufficient impact on the model output, a phenomenon known as the dominant gradient phenomenon. They leverage this phenomenon to search for elements with dominant gradients to generate adversarial samples for adversarial attacks.
Qu et al.~\cite{qu2022synthesising} introduce an adversarial attack algorithm based on the synthetic strategy, which utilizes a conditional variational auto-encoder as a speech synthesizer. Concurrently, an adaptive sign gradient descent algorithm is proposed to enhance the quality of the synthesized samples.

\begin{table}[htb]
\centering
\begin{tcolorbox}
{\color{blue}{\bf Issues to Consider: }The research discussed highlights the potential of leveraging gradient information to guide the generation of audio-visual perturbations, which enables the creation of more potent adversarial examples. Nevertheless, the deployment of algorithms that depend on accessing model gradients presents substantial challenges within black-box scenarios, including significant computational resource consumption.}
\end{tcolorbox}
\end{table}
\vspace{-5mm}
\subsection{Optimization}
By leveraging optimization algorithms to calculate perturbations added to the audio, which minimize perceptibility while maximizing the misleading effect.
Carlini et al.~\cite{carlini2018audio} propose an adversarial attack method that can convert any audio waveform into a specified text. This method optimizes the waveform directly in the white-box setting, enabling speech recognition models to process any audio into the designated target text.
Zheng et al.~\cite{zheng2021black} design a black-box adversarial computation that relies solely on the final decision, without the need for prediction or confidence score information, targeting commercial speech platforms. This algorithm formulates the generation of adversarial examples as a discontinuous large-scale global optimization problem and optimizes it by adaptively decomposing it into sub-problems.

Zhang et al.~\cite{zhang2021attack} design a new adversarial attack strategy that plays carefully crafted adversarial perturbations as a separate audio source. This causes the speaker verification system to misidentify the adversary as the target speaker. The strategy leverages a two-step algorithm to optimize generic adversarial perturbations, making them independent of text content and minimally impacting the recognition of authentication text.
Ko et al.~\cite{ko2023multi} introduce a novel method for creating multi-targeted audio adversarial examples, which are designed to deceive multiple audio models into making incorrect classifications. This method achieves this by finely tuning the adversarial noise added to the original audio samples and specifically adjusting the loss function to maximize the likelihood of each model misclassifying the audio into a predetermined label.
To develop more effective adversarial attack algorithms, Li et al.~\cite{li2023towards} propose a two-step method. First, perturbations are generated for each individual target audio sample to ensure their effectiveness at inducing errors in adversarial samples. Next, these perturbations are aggregated and fine-tuned to form a universal perturbation, creating more efficient adversarial samples.
Wu et al.~\cite{wu2023kenku} present KENKU, an efficient and stealthy black-box adversarial attack framework against automatic speech recognition systems. This framework supports hidden voice commands and integrated command attacks. KENKU optimizes perturbations by leveraging acoustic feature loss and perturbation loss based on Mel-frequency Cepstral Coefficients, avoiding the need for training auxiliary models or estimating gradients.

\begin{table}[htb]
\centering
\begin{tcolorbox}
{\color{blue}{\bf Reflections and Challenges: }The research discussed underscores the application of optimization algorithms in the generation of adversarial examples, which leverage minimizing perceptibility and maximizing misleading effects to generate perturbations. However, how adversarial attacks based on optimization algorithms maintain stealthiness and sustainability in practical applications, especially in complex and variable model environments, warrants further exploration.}
\end{tcolorbox}
\end{table}

\vspace{-4mm}
\subsection{Efficiency}
By measuring the time and number of queries required to generate adversarial examples, the attack efficiency is maximized.
To reduce the consumption of computational resources, Chang et al.~\cite{chang2020audio} introduce an efficient adversarial attack method. They combine pre-training and fine-tuning of the RNN model to accelerate parameter optimization for crafting imperceptible perturbations, constructing adversarial samples. This method is over 400 times faster than the C\&W attack, demonstrating significantly enhanced computational speed.
Wang et al.~\cite{wang2022phonemic} propose a phonemic adversarial attack method that leverages the phoneme density balanced sampling strategy to reduce dependence on training data. Additionally, they use an asynchronous method to optimize phoneme noise, achieving outstanding attack effectiveness and generation speed.
Mun et al.~\cite{mun2022black} introduce an adversarial attack algorithm based on particle swarm optimization in the black-box setting. They leverage adversarial candidates as particles to generate adversarial samples through iterative optimization, which significantly reduces the number of queries and minimizes the risk of detection.

To generate adversarial samples more effectively, Tong et al.~\cite{tong2023query} propose a new gradient estimation strategy, which is named T-NES. This strategy utilizes the inherent temporal correlations in audio samples to accelerate gradient estimation based on probability scores returned by the target model, thereby reducing the number of queries.
Guo et al.~\cite{guo2023phantomsound} introduce a query-efficient black-box adversarial attack toward voice assistants, named PhantomSound, which leverages phoneme-level perturbations to efficiently generate adversarial samples. Specifically, the attacker breaks down target commands into phonemes and gradually injects them into benign audio to create adversarial conditions. This algorithm also reduces the number of queries by optimizing the gradient estimation. Compared to existing methods, this strategy significantly reduces the number of queries required.
Traditional adversarial attack algorithms rely on multiple queries, which is impractical. To address this issue, Fang et al.~\cite{fang2024zero} propose a Zero-Query Adversarial Attack method, named ZQ-Attack. ZQ-Attack initializes adversarial perturbations with a scaled target command audio, which makes the perturbations both covert and efficient. Simultaneously, ZQ-Attack leverages a sequential ensemble optimization algorithm to enhance the transferability of the perturbations. 

\begin{table}[htb]
\centering
\begin{tcolorbox}
{\color{blue}{\bf Insights: }The research discussed highlights the optimization of adversarial attacks by reducing the time and number of queries required to generate adversarial examples. Although these methods have significantly improved efficiency, it remains a challenge whether these efficiency gains consistently hold across different multimodal models and environments.}
\end{tcolorbox}
\end{table}
\vspace{-5mm}
\subsection{Transferability}
Measuring the effectiveness of adversarial attacks involves not only focusing on the success rate but also ensuring outstanding transferability.
Chen et al.~\cite{chen2022push} propose a physical adversarial attack method called PhyTalker. Specifically, PhyTalker generates perturbations at the subphoneme-level, which are optimized and injected into the target speech signal. It compensates for distortions caused by devices and environments through channel enhancement, and uses model ensemble to improve the transferability of the perturbations.
Qi et al.~\cite{qi2023transaudio} introduce a two-stage adversarial attack framework, which features high transferability. In the first stage, an adversarial sample fragment for the target word is generated via text-to-speech models. In the second stage, this adversarial sample is optimized on entire sequences. Additionally, to mitigate adversarial example overfitting to the surrogate model, they also design a score-matching optimization strategy to regularize the training process. Experimental results validate the effectiveness of the proposed algorithm.
Ge et al.~\cite{ge2023advddos} introduce a zero-query adversarial perturbation algorithm, named AdvDDoS, which requires no queries. Specifically, this algorithm includes a popular feature extractor and a local automatic speech recognition model by reversing the robust target-category features, which helps to enhance the transferability of the perturbations. Experimental results show that this algorithm achieves outstanding performance.
Kim et al.~\cite{kim2023generating} investigate the transferability of audio adversarial examples across different model architectures and conditions, and discover that the factors influencing transferability are related to noise sensitivity. Based on these findings, they introduce a new adversarial attack method, which generates highly transferable audio adversarial examples by injecting additive noise during the gradient ascent process.
Farooq et al.~\cite{farooq2025transferable} introduce a transferable GAN-based adversarial attack framework that incorporates a self-supervised audio model to ensure transcription and perceptual integrity, thereby generating high-quality adversarial samples that are more aligned with real-world scenarios.

\begin{table}[htb]
\centering
\begin{tcolorbox}
{\color{blue}{\bf Issues to Consider: }The research discussed underscores how multiple methods enhance the attack success rate while simultaneously improving the transferability of perturbations. It is noteworthy that these complex attack frameworks may require additional computational resources and time, which are important for the practicality of the attacks.}
\end{tcolorbox}
\end{table}

\vspace{-5mm}
\subsection{Stealthiness}
The stealthiness of perturbations is a necessary factor to ensure the successful implementation of adversarial attacks.
Gong et al.~\cite{gong2019audidos} introduce an audio adversarial attack algorithm, named a denial-of-service attack. This algorithm trains a universal adversarial perturbation to maximize the misclassification rate while limiting the perturbation's amplitude to minimize the perceptibility of the attack. Notably, this attack can be implemented in real-time and over-the-air during user interactions with voice control systems, and it is unaffected by user commands or interaction times.
Liu et al.~\cite{liu2020weighted} introduce audio adversarial examples with weighted sampling, a method that considers the numbers and weight of distortion to enhance the effectiveness of adversarial attacks. Furthermore, to improve the stealthiness of the perturbations, they also introduce a denoising algorithm during the training process. This strategy makes the audio adversarial attacks not only harder to detect but also maintains the effectiveness and precision of the attacks.
Li et al.~\cite{li2020practical} propose an adversarial attack algorithm targeting speaker recognition systems. Specifically, they utilize gradient-based adversarial machine learning algorithms to generate adversarial examples that include well-crafted inconspicuous noise, deceiving the speaker recognition system into making incorrect predictions.
Du et al.~\cite{du2021robust} introduce a novel adversarial attack framework, which includes components for physical sample simulation (PSS) and adversarial patch noise generation (APNG). The PSS component is used to simulate real-audio with selected room impulse responses for training the adversarial patches. The APNG component uses the voice activity detector to generate imperceptible audio adversarial patch examples.
Qiu et al.~\cite{qiu2024boosting} introduce a frequency-weighted perturbation algorithm for adversarial attacks based on environmental sounds. This algorithm integrates auditory thresholds with psychoacoustic principles to generate adversarial samples that are difficult to detect.

\begin{table}[htb]
\centering
\begin{tcolorbox}
{\color{blue}{\bf Issues to Consider: }The research presented delves into the study of stealth in adversarial attacks, focusing on generating perturbations that are difficult to detect while ensuring the effectiveness of the attack. It is worth mentioning that in complex audio-visual environments, maintaining stealth becomes more challenging and may require higher computational costs. }
\end{tcolorbox}
\end{table}

\vspace{-5mm}
\subsection{Semantic}
Leveraging semantic implementation for adversarial attacks is indeed an effective strategy, which involves adding targeted but semantically consistent perturbations.
Yu et al.~\cite{yu2023smack} propose a semantically meaningful audio adversarial attack algorithm, which leverages semantic perturbations to modify the inherent speech attributes. To construct adversarial samples, they design an adapted generative model that enables fine-grained control of prosody. The model ensures that the modified samples still semantically represent the same speech and preserve the speech quality.
Dou et al.~\cite{dou2023adversarial} formulate adversarial attacks as an optimization problem, aiming to minimize the angular deviation between the embeddings of the transformed input and the perturbed audio. This method can effectively implement adversarial attacks in the multi-modal setting, which avoids the discrepancies between different modalities.
Cheng et al.~\cite{cheng2024alif} introduce a black-box adversarial attack algorithm based on linguistic features, which leverages the reciprocal process of text-to-speech and automatic speech recognition models to generate perturbations in the linguistic embedding space. This algorithm can generate adversarial samples with only a single query, demonstrating high efficiency.

\begin{table}[htb]
\centering
\begin{tcolorbox}
{\color{blue}{\bf Implications: }The research discussed highlights the effectiveness of introducing semantically consistent perturbations in adversarial attacks, enhancing the stealthiness and naturalness of adversarial samples. Moreover, this strategy is also applicable to research on backdoor attacks targeting audio-visual tasks.}
\end{tcolorbox}
\end{table}
\vspace{-5mm}
\subsection{Advanced models}
Research on adversarial attacks targeting multimodal large language models (MLLMs) deserves more attention.
Yang et al.~\cite{yang2024can} build the evaluation benchmark for chat-audio MLLMs, which includes four distinct types of adversarial audio attacks to comprehensively assess the resilience of MLLMs. In this benchmark, they use Standard, GPT-4-based, and Human Evaluation methods to assess the robustness of MLLMs.
Huang et al.~\cite{huang2025image} explore the transferability of adversarial video samples across video-based MLLMs. They introduce a highly transferable attack method, which leverages an image-based multimodal model as the surrogate model to craft adversarial video samples. Experimental results show that this method effectively disrupts various video-based MLLMs.
To explore the security of video-based MLLMs against adversarial attacks, Li et al.~\cite{li2024fmm} introduce an adversarial attack algorithm tailored for video-based MLLMs. The core of this algorithm involves generating multimodal, flow-based adversarial perturbations on a few frames of the video, misleading video-based MLLMs into generating incorrect responses. The algorithm combines video features and textual features for the attack, exploiting the transferability of features across different modalities to breach the model's security.

\begin{table}[htb]
\centering
\begin{tcolorbox}
{\color{blue}{\bf Implications: }Although previous adversarial attack algorithms have a certain level of transferability, simple perturbations often fail against MLLMs due to their inherent protective mechanisms. Therefore, exploring more effective perturbation strategies for MLLMs is particularly important. This research direction not only helps understand the vulnerabilities of MLLMs but also promotes the development of more robust defense mechanisms.}
\end{tcolorbox}
\end{table}

\vspace{-5mm}
\subsection{Evaluation}
Vadillo et al.~\cite{vadillo2022human} measure the extent of distortion in audio adversarial examples, emphasizing that high-quality audio adversarial examples must maintain their adversarial nature while avoiding human detection. They assess these examples using an analytical framework based on various factors, demonstrating that the conventionally used metrics are not reliable measures of perceptual similarity in the audio domain for adversarial examples.

\section{Backdoor Attack}\label{backdoor}
For audio-visual-based multimodal models, the motivation behind backdoor attacks is to embed malicious triggers into the target model, thereby manipulating the model's response. In this section, we categorize backdoor attacks targeting audio-visual models from two perspectives: malicious example design and victim model fine-tuning, as shown in Table \ref{table_backdoor}.

\begin{table*}[t]
\centering
 \caption{Summary of backdoor attack methods against audio-visual-based multimodal models, which includes the method name, capability, attack characteristics, representative models and tasks, and partial contributions.}
\fontsize{6.2pt}{10pt}\selectfont
\selectfont 
\renewcommand{\arraystretch}{1.0}
\resizebox{\textwidth}{!}{
\begin{tabular}{rccccc}
\hline
Method           &Capability  &Characteristics     & Model &  Task & Contribution\\
\hline
S\&DBA~\cite{al2024look}         &White-box        &Triggers    &CNN    &Video Action Recognition  & First to investigate audiovisual backdoor attacks in video models\\
Aliasing~\cite{lee2023aliasing}  &White-box        &Triggers    &Vision Transformers   &Speech Recognition  &Exploits strided layers for effective and stealthy backdoor attacks.\\
PALETTE~\cite{ze2023ultrabd}     &Black-box        &Triggers    &Inflated 3D ConvNet    &Video Action Recognition  & First physically-realizable video backdoor attack framework\\
TrojanRoom~\cite{chen2024devil}	 &White-box       &Triggers    &BC-ResNet    &Speaker Recognition  &Introduces RIR-based physical triggers for audio backdoor attacks\\
FlowMur~\cite{lan2024flowmur}          &Gray-box &Optimization &DNN &Speech Recognition &Proposes a stealthy audio backdoor attack with limited knowledge requirement \\
SilentTrig~\cite{tang2024silenttrig}   &Gray-box &Optimization &DNN &Speaker Identification &Introduces imperceptible backdoor attack using audio steganography \\
SMA~\cite{zheng2023silent}             &Gray-box &Stealthiness &LSTM  &Speech Recognition &Inaudible backdoor attack exploiting microphone nonlinearity \\
TrojanModel~\cite{zong2023trojanmodel} &Gray-box &Stealthiness &LSTM &Speech Recognition & Introduces a practical Trojan attack leveraging unsuspicious triggers\\
MagBackdoor~\cite{liu2023magbackdoor}  &Black-box &Stealthiness &Closed-source &Voice Command Injection &Introduces magnetic field attack via loudspeaker-based backdoor \\
IRBA~\cite{yao2025imperceptible}          &Gray-box &Rhythm &DNN  &Speech Recognition &Proposes stealthy rhythm transformation for undetectable backdoor attacks \\
Adv-audio~\cite{kong2019adversarial}      &Gray-box &Adversarial &DNN &Speech Recognition &Proposes imperceptible audio hiding method activating backdoor \\
Fake~\cite{ye2023fake} &Gray-box &Conversion &RawNet3 &Speaker Recognition &Utilizes voice conversion for stealthy backdoor attacks in speech models \\
EmoAttack~\cite{yao2024emoattack} &Gray-box &Conversion &SincNet &Speaker Verification &Utilizes emotional voice conversion for speech backdoor attacks\\
PBSM; VSVC~\cite{cai2024towards}  &Gray-box &Conversion &LSTM &Speech Recognition &Utilizes sound elements for stealthy backdoor attacks \\
JingleBack~\cite{koffas2023going}&White-box &Style &LSTM &Speech Classification &Introduces stylistic triggers using guitar effects for audio backdoors \\
MarketBack~\cite{mengara2024trading}&Gray-box &Style &Transformer &Speech Recognition &Proposes backdoor attack using stochastic investment models\\
PhaseBack~\cite{ye2023stealthy}    &White-box &Spectrogram &DNN &Speaker Recognition &Proposes inaudible phase-injection triggers for backdoor attacks. \\
IBA~\cite{zhang2024inaudible}&Gray-box &Spectrogram &DeepSpeaker &Speaker Recognition &Introduces inaudible frequency-domain triggers for backdoor attacks \\
PAS~\cite{liu2022backdoor}   &Gray-box &Steganography &DNN &Speaker Verification &Introduces personalized audio steganography for backdoor attacks  \\
Echo~\cite{zhang2024audio}    &Gray-box &Steganography &LSTM &Speech Recognition &Proposes echo hiding for stealthy backdoor attacks \\
\hline
FAB~\cite{yunsounding}          &Gray-box &Optimization &Transformer &Speech Recognition &Proposes task-agnostic backdoor attacks on AFMs \\
PIBA~\cite{shi2022audio}        &Gray-box &Optimization &SincNet &Speaker Recognition &Introduces position-independent audio backdoor attack\\
OPP~\cite{liu2022opportunistic} &Gray-box &Optimization &LSTM &Speech Recognition &Proposes audible backdoor attack using daily ambient noise as triggers\\
BAGS~\cite{han2024backdooring}  &White-box &Optimization &OpenVQA &VQA; AVSR &Introduces data-efficient backdoor attack for multimodal learning \\
\hline
\end{tabular}}
\label{table_backdoor}
\end{table*}

\subsection{Malicious example design}
Constructing poisoned samples is an indispensable step in implementing backdoor attacks, which involves the design of the trigger and the selection of the target label~\cite{zhao2024exploring,zhao2025clean}.
Koffas et al.~\cite{koffas2022can} investigate the impact of backdoor attacks on speech recognition systems by embedding inaudible triggers into training samples, demonstrating the system's vulnerabilities. Additionally, they validate the effectiveness of the trigger's duration, position, and type.
Xin et al.~\cite{xin2022natural} leverage sounds that are ordinary in nature or in our daily lives as triggers to implement backdoor attacks. Experimental results show that with only 5\% of poisoned samples, an attack success rate of nearly 100\% can be achieved.
Luo et al.~\cite{luo2022practical} launch backdoor attacks to verify the vulnerability of speaker recognition systems in both digital and physical spaces. Specifically, they examine the effects of the trigger's position, intensity, length, frequency characteristics, and the poison rate of the poisoned samples on the success rate of the backdoor attacks.
Bartolini et al.~\cite{bartolini2024hidden} introduce a new backdoor attack algorithm, which maps different environmental trigger sounds to target phrases of various lengths during the model fine-tuning phase, demonstrating the potential security vulnerabilities of the Whisper model. 
Guo et al.~\cite{guo2023masterkey} investigate the limitations of existing backdoor attacks and design a universal backdoor capable of attacking arbitrary targets. Their algorithm injects poisoned audio samples into the training data, embedding the backdoor during the fine-tuning of the target model. Attackers can then trigger the backdoor by playing specific audio, thus executing the attack without altering legitimate user data.
Orson Mengara~\cite{mengara2024backdoor} introduces a dynamic label inversion backdoor attack algorithm, which leverages clapping as an audio trigger while maintaining the correctness of the labels for the poisoned samples.
Al Kader Hammoud et al.~\cite{al2024look} revisit the traditional backdoor attacks in the image domain and expand them to the audio domain in two ways: statically and dynamically, leading to a highly effective attack success rate. Additionally, they also explore multi-modal backdoor attacks against video action recognition models.

Furthermore, Ze et al.~\cite{ze2023ultrabd} also explore the use of adversarial ultrasound as the triggers for implementing backdoor attacks. They introduce randomness in the synchronous time and the relative amplitude ratio between the adversarial ultrasound and the legitimate user to enhance the effectiveness of the backdoor attack.
Gong et al.~\cite{gong2023palette} present a physically-realizable backdoor attack algorithm, named PALETTE, which features two special design choices. First, they utilize natural-light-alike RGB offsets as triggers without the need to modify video files. Second, they leverage rolling operations to make the backdoored model more robust to temporal asynchronization. Experimental results show that PALETTE outperforms existing video backdoor attacks.
Wei et al.~\cite{lee2023aliasing} introduce the aliasing backdoor, which explores the implementation of backdoor attacks in pretrained models based on the aliasing effect induced by sampling. Specifically, they introduce aliasing errors in the stride layers of the model, manipulating the input data to produce misleading outputs. This algorithm features the characteristics of being low-cost and data-free.
Liu et al.~\cite{liu2024mtbv} propose a multi-trigger backdoor attack algorithm, which simultaneously maps multiple triggers to target attack objects, effectively enhancing the success rate of the backdoor attacks.
Chen et al.~\cite{chen2024devil} present the TrojanRoom algorithm, which bridges the gap between digital and physical audio backdoor attacks. Specifically, they leverage the room impulse response as a physical trigger to poison target samples without the need to implant additional explicit triggers.

\noindent {\bf Optimization }Considering that traditional backdoor attack algorithms often require significant expertise, which limits their widespread adoption, Lan et al.~\cite{lan2024flowmur} introduce FlowMur. This approach can be launched with limited knowledge and formulates trigger generation as a dynamic optimization problem based on an auxiliary dataset and a surrogate model. Furthermore, they develop an adaptive data poisoning method to optimize the concealment of the trigger.
Li et al.~\cite{li2023enrollment} implement a backdoor attack on automatic speech recognition systems during the enrollment stage via adversarial ultrasound. To enhance real-world robustness, they generate the ultrasonic backdoor by augmenting the optimization process and optimizing the ultrasonic signal using sparse frequency points, precompensation, and single-sideband modulation.
Existing audio-based backdoor attack algorithms often utilize discernible noise as triggers, making them susceptible to detection by defense algorithms. Tang et al.~\cite{tang2024silenttrig} introduce an imperceptible backdoor attack method named SilentTrig. This method embeds the trigger within benign audio samples by leveraging an optimized steganographic network and implementing a two-stage adversarial optimization process to ensure that the poisoned samples are acoustically indistinguishable from the benign samples, significantly increasing the undetectability of the attack.

\noindent {\bf Stealthiness }To further optimize the stealth of audio triggers, Zheng et al.~\cite{zheng2023silent} propose an inaudible grey-box backdoor attack, named SMA. Specifically, they utilize the nonlinear effects of microphones to inject an inaudible ultrasonic trigger into the samples. Additionally, to enhance the robustness and transferability of the trigger in the physical world, an optimization algorithm has been designed. Experimental results indicate that the SMA algorithm achieves a feasible success rate for backdoor attacks.
Zong et al.~\cite{zong2023trojanmodel} consider the use of a piece of background music as an unsuspicious trigger, which avoids detection by defense algorithms. Additionally, they introduce a small network called the TrojanModel to incorporate backdoor features, eliminating the requirement for retraining of the target model to implement the backdoor attack.
Liu et al.~\cite{liu2023magbackdoor} propose MagBackdoor, a backdoor attack algorithm that injects malicious commands via a loudspeaker, compromising the linked voice interaction system. They self-design a prototype that can emit magnetic fields modulated by voice commands, making the attack more covert and applicable to real-world scenarios.
To avoid audible triggers being detected by human ears, Ye et al.~\cite{ye2024breaking} present an inaudible backdoor attack named PaddingBack. They exploit the widely-used speech signal operation, padding, to serve as the backdoor trigger. Experimental results show that PaddingBack achieves a feasible attack success rate and resistance to defense mechanisms.
Xiong et al.~\cite{xiong2024phoneme} argue that the human ear's lower sensitivity to consonant phonemes can be exploited to implement backdoor attacks. Therefore, they introduce a backdoor attack method based on phoneme substitution. Specifically, a selection and substitution strategy is designed for triggers, which can covertly replace phonemes to construct poisoned samples.

\noindent {\bf Rhythm transformation }To enhance the stealthiness of poisoned samples, Yao et al.~\cite{yao2025imperceptible} introduce a non-neural and efficient backdoor attack algorithm named RSRT. Specifically, the RSRT algorithm designs a trigger that stretches or squeezes the mel spectrograms and then recovers them back to signals. This operation helps maintain the timbre and content unchanged, thereby increasing the stealthiness of the poisoned samples.

\noindent {\bf Adversarial perturbation }Kong et al.~\cite{kong2019adversarial} introduce a new audio information hiding algorithm that combines high hiding capacity with excellent imperceptibility. Specifically, the algorithm embeds hidden information, which can be used as a trigger, into the audio signal to create stego audio. This stego audio is then used to train and implement backdoor attacks, posing a serious threat to automatic speech recognition systems.

\noindent {\bf Voice conversion }To enhance the concealment of triggers, Cai et al.~\cite{cai2022pbsm} adopt a strategy that combines pitch boosting with sound masking to optimize the trigger, reducing its detectability by human ears. Experimental results demonstrate that this algorithm can achieve an attack success rate of over 90\% with just a 1\% poisoning rate.
Additionally, Cai et al.~\cite{cai2022vsvc} explore the backdoor attack algorithm based on voice conversion to enable multi-target attacks. This algorithm leverages voice conversion to transform the timbre of speech, evading detection by human ears and enhancing the stealthiness of the backdoor attacks.
Ye et al.~\cite{ye2023fake} present a sophisticated backdoor attack algorithm that leverages sample-specific triggers generated through voice conversion. This method specifically utilizes a pre-trained voice conversion model to create these triggers, ensuring that the altered audio samples remain devoid of any extraneous audible noise. This approach not only preserves the naturalness of the audio but also embeds the malicious trigger with enhanced subtlety and effectiveness.
Yao et al.~\cite{yao2024emoattack} consider emotion as a higher-level subjective perceptual attribute inherent in speech, which can serve as triggers for backdoor attacks. Therefore, they design an emotional voice conversion algorithm to generate high-quality triggers. Experimental results indicate that using emotion as triggers can achieve a feasible success rate for backdoor attacks.
Schoof et al.~\cite{schoof2024emoback} leverage the emotional prosody of speakers to construct dynamic, inconspicuous triggers, thereby enhancing the stealthiness of backdoor attacks.
Cai et al.~\cite{cai2024towards} explore elements of sound, such as pitch and timbre, as triggers to enhance the stealthiness of backdoor attacks. Specifically, they design two distinct backdoor attack algorithms. Firstly, they utilize a short-duration, high-pitched signal as the trigger, which is implanted into the target sample. Subsequently, they increase the pitch of the remaining audio clips to ``mask" this trigger, thereby avoiding detection by defense algorithms. Secondly, they manipulate the timbral characteristics of the victim audio samples using a voiceprint selection module to generate poisoned samples.

\noindent {\bf Style }Koffas et al.~\cite{koffas2023going} explore stylistic-based audio backdoor attack triggers, which demonstrate the effectiveness of six different stylistic triggers on audio models.
Orson Mengara~\cite{mengara2024trading} manipulates the stylistic properties of audio as triggers for backdoor attacks, utilizing stochastic investment models coupled with Bayesian methodologies to generate the poisoned samples. Furthermore, to enhance the stealthiness of backdoor attacks, they ensure that the labels of the poisoned samples remain unchanged, a strategy known as clean-label backdoor attacks. Furthermore, Orson Mengara~\cite{mengara2024art} introduces a dynamic trigger strategy to reduce the distinguishability between poisoned samples and clean samples. They insert triggers into the audio signal by leveraging the short-time Fourier transform to acquire the speech spectrogram, constructing the poisoned samples.

\noindent {\bf Audio Spectrogram }Ye et al.~\cite{ye2023stealthy} present a novel phase-injection backdoor attack algorithm that utilizes the Short-Time Fourier Transform (STFT) to process audio samples. Specifically, the spectrogram of the target sample is obtained leveraging the STFT algorithm, then split into the phase and amplitude spectra. Subsequently, a malicious trigger is implanted into the phase spectrum. Finally, the inverse STFT is used to obtain the poisoned samples. This algorithm effectively reduces the audibility of the trigger, enhancing its stealthiness.
Zhang et al.~\cite{zhang2024inaudible} propose a novel audio backdoor attack that utilizes inaudible triggers within the frequency domain of audio spectrograms. Specifically, they develop  a strategy that automatically generates inaudible triggers in the supported spectrum and optimizes the robustness of these triggers.

\noindent {\bf Audio Steganography }Liu et al.~\cite{liu2022backdoor} introduce a novel algorithm for an audio steganography-based personalized trigger backdoor attack. Specifically, a pre-trained audio steganography network is employed to implicitly write personalized information into the target samples, which can effectively enhance the concealment of the attack. Moreover, the algorithm modifies only the frequency and pitch of the target samples, making it more difficult to detect.
Zhang et al.~\cite{zhang2024audio} introduce a backdoor attack method embedded in the frequency domain based on echo hiding, a technique in audio steganography that embeds hidden information into the frequency spectrum. This approach effectively avoids detection of the trigger while ensuring the audio quality is maintained.

\noindent {\bf Federated learning }Wu et al.~\cite{wu2024web} propose a backdoor attack algorithm designed for federated learning-based automatic speaker verification systems. Specifically, contend that the complexity of voiceprint data facilitate the introduction of subtle perturbations, allowing for the embedding of poisoned samples with triggers into the training dataset to execute backdoor attacks. To augment the stealthiness of these attacks, they incorporate triggers into utterances in a manner that is more inconspicuous.
Xu et al.~\cite{xu2025sample} propose a novel backdoor attack in federated learning, named GhostB. This algorithm leverages the behavior of neurons that produce specific values to serve as triggers. It avoids modifying training samples and does not rely on dropout, effectively enhancing both the stealth and efficiency of the backdoor attack.

\begin{table}[htb]
\centering
\begin{tcolorbox}
{\color{blue}{\bf Critical Points: } The research discussed highlights the leverage of stealthiness and natural triggers, such as inaudible sounds or common everyday noises, which help to enhance the stealth and attack success rate. Additionally, optimized dynamic triggers that can adapt to various environments have enhanced the robustness and transferability of attacks. However, these triggers are designed to operate under specific conditions, which limits their applicability. Furthermore, as defense strategies continue to improve, these triggers may become easier to detect, reducing the effectiveness of backdoor attacks.}
\end{tcolorbox}
\end{table}

\vspace{-6mm}
\subsection{Victim model fine-tuning}
To more effectively implement backdoor attacks, some research assumes that attackers may manipulate the training process.
Yun et al.~\cite{yunsounding} present the FAB algorithm, specifically designed to target acoustic foundation models. This algorithm employs physically plausible and unobtrusive auditory triggers, such as the sounds of dog barks and alarms. By optimizing the training loss, it ensures that the model generates irrelevant representations when the input contains these triggers, while maintaining normal performance when no triggers are present. Shi et al.~\cite{shi2022audio} explore backdoor attacks in the audio domain, demonstrating that an unnoticeable audio trigger can easily launch such attacks. They jointly optimize the trigger and the victim model during the training phase to construct an unnoticeable and position-independent audio trigger. Specifically, they optimize the trigger's position to ensure it is position-independent, and design an algorithm that mimics environmental sounds to make the trigger resemble something unnoticeable. Given the susceptibility of conventional audio triggers to preprocessing, Liu et al.~\cite{liu2022opportunistic} introduce an audible backdoor attack algorithm tailored for speech recognition systems. Specifically, the triggers employed are ambient noises from daily contexts, capitalizing on people's auditory inertia, making these triggers naturally stealthy and easily overlooked. Simultaneously, they introduce a dual-adaptive backdoor augmentation method to enhance the efficiency of the backdoor attack, which can facilitate robust model poisoning and achieve a high attack success rate. 

To explore the effectiveness of backdoor attacks in multimodal learning, Han et al.~\cite{han2024backdooring} propose a data and computation efficient algorithm for backdoor attacks. Initially, they design a backdoor gradient-based score to quantify the contribution of each data sample to the backdoor attack during the training phase. Subsequently, a searching strategy is introduced to efficiently determine the optimal poisoning modalities and data samples. Extensive experiments validate the effectiveness and efficiency of the algorithm.
To address the problem of traditional data poisoning backdoor attacks being easily detected during both the training and inference stages, Xiao et al.~\cite{xiao2024phoneme} introduce a phoneme mixing and multitask learning algorithm to implement backdoor attacks without altering the input samples. Specifically, the attackers utilize certain phonemes as semantic triggers, while simultaneously leveraging a multiple gradient descent algorithm to optimize multitask learning, enhancing both the effectiveness of the backdoor attacks and the accuracy of classification. 

\begin{table}[htb]
\centering
\begin{tcolorbox}
{\color{blue}{\bf Challenges: }As another approach to backdoor attacks, manipulating the model training process often achieves a more effective success rate. However, algorithms that depend on accessing the model training process are difficult to implement in the black-box setting and may also require more training iterations.}
\end{tcolorbox}
\end{table}

\vspace{-5mm}
\section{Jailbreak Attack}\label{jailbreak}
Although previous research demonstrates the vulnerability of LLMs or MLLMs to jailbreak attacks~\cite{xu2024comprehensive,liu2024jailbreak}, exploring jailbreaking attacks related to audio-visual is equally worthy of attention.
In this section, we present the jailbreaking attacks to audio-visual-based MLLMs with three types: prompt engineering, signal perturbation, and multimodal alignment.
We provide a summary as shown in Table \ref{table_jailbreak}.

\begin{table*}[t]
\centering
 \caption{Summary of jailbreak attack methods against audio-visual-based multimodal models, which includes the method name, capability, attack characteristics, representative models and tasks, and partial contributions.}
\fontsize{6.2pt}{10pt}\selectfont
\selectfont 
\renewcommand{\arraystretch}{1.0}
\resizebox{\textwidth}{!}{
\begin{tabular}{rccccc}
\hline
Method           &Capability  &Characteristics     & Model &  Task & Contribution\\
\hline
JMLLM~\cite{mao2024divide} &Gray-box  &Hybrid  & GPT-4o & TriJail &Achieved high attack success using multimodal hybrid strategy\\
BoN~\cite{hughes2024best}      &Black-box  &Prompt  & GPT-4  & HarmBench  &Proposes jailbreak attack combined with augmentation strategies  \\
\hline
Audio-j~\cite{gupta2025bad} &White-box  &Perturbation  &SALMON-N 7B  & Jailbreak audio &Demonstrated audio jailbreaks bypassing ALM alignment mechanisms effectively\\
AdvWave~\cite{kang2024advwave} &White-box  &Adversarial  &Llama-Omni  &AdvBench-Audio  &AdvWave achieves high success rates while maintaining audio stealthiness  \\
SS-Jail~\cite{yang2024audio} &Black-box  &Evaluation  & Qwen2-Audio &AdvBench  & Revealed audio LMM vulnerabilities via speech-specific jailbreak \\
AET; EADs~\cite{xiao2025tune} &Black-box  &Editing  &Qwen2-Audio  &AdvBench  &Introduced AET and EADs to assess LALMs' security vulnerabilities  \\
\hline
VideoJail~\cite{hu2025videojail} &Black-box  &Embedding  &Qwen2-VL  &HADES  &Video modality increases security risks in multimodal models  \\
T2VSafetyBench~\cite{miao2024t2vsafetybench} &Black-box  &Benchmark  & GPT-4 & VidProM  & Evaluated video generation safety across twelve critical aspects \\
Flanking~\cite{chiu2025say} &Black-box  &Prompt  &Gemini  & Jailbreak audio &  Proposes Flanking Attack, bypassing defenses with narrative-driven prompts\\
VOICEJAILBREAK~\cite{shen2024voice}  &Black-box  &Elements  &GPT-4o  &ForbiddenQuestionSet   &Enhanced voice jailbreak attack effectiveness using fictional storytelling elements.  \\
\hline
\end{tabular}}
\label{table_jailbreak}
\end{table*}

\subsection{Prompt Engineering}
Mao et al.~\cite{mao2024divide} introduce a hybrid strategy-based multimodal jailbreak attack algorithm. Specifically, they combine a series of strategies including alternating translation, word encryption, harmful injection, and feature collapse, which ensure that the prompts remain adversarial while bypassing the model's defense mechanisms. Additionally, they set up the hybrid strategy with both single-query and multi-query options. Compared to single-query, although multi-query requires more time, it helps enhance jailbreak performance.
Hughes et al.~\cite{hughes2024best} propose the Best-of-N Jailbreaking, a black-box jailbreaking attack algorithm to extract harmful information from audio-visual-based MLLMs. Specifically, they repeatedly sample variations of a prompt with a combination of augmentation strategies, such as speed, pitch adjustments, and background noise adjustments, to induce MLLMs to produce unsafe outputs.

\subsection{Signal Perturbation}
Gupta et al.~\cite{gupta2025bad} demonstrate universal jailbreaks in the audio modality for MLLMs by constructing adversarial perturbations. They also provide an insightful analysis showing that imperceptible, first-person toxic speech is the most effective type of perturbation for eliciting toxic outputs. Specifically, these perturbations embed linguistic features within the audio signal.
Kang et al.~\cite{kang2024advwave} propose a jailbreak framework against MLLMs, named AdvWave. The AdvWave framework leverages a dual-phase optimization method to address the gradient shattering problem. Additionally, they also introduce an adaptive adversarial target search algorithm to dynamically construct adversarial queries targeted at specific responses.
Yang et al.~\cite{yang2024audio} evaluate the security of five audio-based MLLMs against various queries: (i) harmful questions in both audio and text formats, (ii) harmful questions in text format accompanied by distracting non-speech audio, and (iii) speech-specific jailbreaks. Furthermore, they also introduce a speech-specific jailbreaking method, which decomposes harmful words into letters to conceal them in the audio input, thereby bypassing the model's security protections.
Xiao et al.~\cite{xiao2025tune} explore audio-specific edits to implement jailbreak attacks on MLLMs and introduce the Audio Editing Toolbox (AET). The AET offers a range of editing tools for audio tone adjustment, word emphasis, and noise injection. It includes edited audio datasets, which can serve as benchmark datasets for testing jailbreak attacks on MLLMs.

\subsection{Multimodal Alignment}
To verify the security related to modality alignment, Hu et al.~\cite{hu2025videojail} introduce a novel jailbreak attack algorithm, named VideoJail, which induces video-based MLLMs to amplify harmful content in images. Specifically, by leveraging carefully crafted text prompts, VideoJail can induce the model to focus its attention on malicious queries, successfully breaching the model's security mechanisms.
Miao et al.~\cite{miao2024t2vsafetybench} construct a comprehensive benchmark to evaluate the security of text-to-video models, which employ jailbreak attack-based prompts. Experimental results indicate that no single video-based MLLM excels across all critical aspects, necessitating a further trade-off between the usability and safety of the video model.
Chiu et al.~\cite{chiu2025say} leverage audio samples to circumvent the constraints of traditional text, breaking through the defense mechanisms of MLLMs. Specifically, they embed harmful prompts within seemingly harmless ones, reducing the MLLMs' ability to recognize and block harmful content by increasing situational complexity and introducing ambiguity. At the same time, they construct a sequence of multiple queries, placing the critical adversarial query in the middle of the sequence to reduce the risk of triggering safety mechanisms.
Shen et al.~\cite{shen2024voice} systematically measure the security of the GPT-4o model against jailbreak attacks. The results indicate that the GPT-4o model exhibits good resistance to malicious queries when leveraging voice mode. Furthermore, they also present a novel voice jailbreak attack algorithm that attempts to persuade the GPT-4o model through fictional storytelling. Extensive experiments validate the feasibility of this attack method.

\subsection{Evaluation }
Ying et al.~\cite{ying2024unveiling} evaluate the capabilities of GPT-4o against jailbreak attacks across three modalities: text, speech, and image. Their findings expose a new attack surface for jailbreak attacks on GPT-4o specifically in the speech modality.

\begin{table}[htb]
\centering
\begin{tcolorbox}
{\color{blue}{\bf Reflections and Challenges: }The research discussed underscores that the inherent defense mechanisms of state-of-the-art MLLMs can be circumvented by malicious audio manipulations and queries, thereby producing harmful outputs. However, there is less research on jailbreaking attacks targeting audio-visual-based MLLMs, which requires more attention. Additionally, due to variations in model structures, jailbreaking attack algorithms may demonstrate varying performance and require substantial computational resources.}
\end{tcolorbox}
\end{table}
\vspace{-6mm}
\section{Other attack}\label{other attack}
Roy et al.~\cite{roy2017backdoor} leverage the non-linear diaphragm and amplifiers of microphones to create ``shadows" that facilitate covert acoustic communication. This method is applied in acoustic denial-of-service attacks. Li et al.~\cite{li2024safeear} propose an algorithm for detecting audio deepfakes that operates without relying on semantic information. Specifically, the algorithm decouples semantic and acoustic information in the audio, utilizing only the acoustic information for detection. This approach effectively identifies a variety of deepfake technologies while maintaining a low error rate.
\vspace{-3mm}
\section{Defense}\label{defense}
Compared to research on attack algorithms targeting audio-visual-based multimodal models, exploring defense algorithms appears to be more crucial~\cite{zhao2024defending}. Therefore, we discuss the algorithms for defending against adversarial and backdoor attacks from three perspectives: sample detection, optimization, and purification, as shown in Table \ref{table_defense}. It is worth noting that there are no defense algorithms available for jailbreak attacks, especially for audio-visual-based multimodal models.

\begin{table*}[t]
\centering
 \caption{Summary of defense algorithms against audio-visual attacks, which includes the method name, capability, defense characteristics, representative models and tasks, and partial contributions.}
\fontsize{6.2pt}{10pt}\selectfont
\selectfont 
\renewcommand{\arraystretch}{1.0}
\resizebox{\textwidth}{!}{
\begin{tabular}{rccccc}
\hline
Method           &Capability  &Characteristics     & Model &  Task & Contribution\\
\hline 
WaveGuard~\cite{du2020unified} &White/Gray-box  &Sample detecting   & Deepspeech  &  Speech Recognition  & Detects audio adversarial examples without model retraining \\
CS-Detect~\cite{kwon2019poster}&White/Gray-box   &Sample detecting   & RNN   & Speech Recognition  & Detects audio adversarial examples using classification score patterns \\
Agitated~\cite{park2024toward}&White/Black-box  &Sample detecting   & RNN  &  Speech Recognition & Introduces noise to logits for adversarial example detection \\
Robust~\cite{esmaeilpour2019robust}&White/Black-box   &Optimization   & SVM   & Audio Classification  & Uses DWT and SVM for robust audio attack defense \\
SNR~\cite{lu2023adversarial}&White-box  &Optimization   & ResNet18  &  Audio Classification & Applies SNR, APGD attacks, and Cutmix for robustness improvement \\
HF-Smoothing~\cite{olivier2021high}&White-box  &Purification   & CNN  &  Speaker Identification &Utilizes high-frequency smoothing to enhance adversarial robustness\\
FeCo~\cite{chen2022towards}&White/Black-box  &Purification   &  CNN &  Speaker Recognition & Combines feature compression with adversarial training for robust defenses \\
AudioPure~\cite{wu2023defending}&White/Black-box  &Purification   & Diffusion  &Speech Command Recognition   &Uses diffusion models for effective adversarial audio purification  \\
\hline
SpeechGuard~\cite{xin2024speechguard}&White/Black-box &Sample detecting &Autoencoder &Speech Recognition &Combines detection and purification for robust backdoor attack defense \\
Sniper~\cite{guo2023masterkey}&Black-box &Sample detecting & t-SNE &Speaker Verification & Effectively cleanses datasets using sniper-based defense mechanism\\
GN-FT~\cite{zhou2025gradient}&White-box  &Optimization &LSTM &Speech Recognition &Reduces backdoored effects by penalizing high-gradient neurons \\
Silero-VAD~\cite{bartolini2024hidden}&White-box &Purification &Transformer &Speech Recognition &Uses VAD to filter malicious triggers and protect models \\
KDDF~\cite{chen2023knowledge} &White-box&Purification & ResNeXt &Speech Recognition & Detects triggers and recovers poisoned audio data using KD.\\
\hline
\end{tabular}}
\label{table_defense}
\end{table*}

\subsection{Adversarial attack}
\noindent {\bf Sample detecting }To detect adaptive adversarial audio samples, Du et al.~\cite{du2020unified} introduce a unified detection framework, which combines noise padding and sound reverberation. Specifically, an adaptive artificial speech generator is designed to enhance the effectiveness of detection. Additionally, they also design multiple noise padding strategies to disrupt the continuity of adversarial noise. The advantage of this framework is that it is applicable to various automatic speech recognition systems without the need for additional training.
Kwon et al.~\cite{kwon2019poster} propose an adversarial audio sample detection algorithm, which adds a new low-level distortion to the samples under inspection. If the classification result is sensitive, then the sample is identified as an adversarial sample; otherwise, it is considered an original sample. Experimental results validate the effectiveness of this algorithm.
Hussain et al.~\cite{hussain2021waveguard} introduce an adversarial sample detection framework named WaveGuard, which incorporates an audio transformation module. The fundamental premise of this module is that adversarial examples induce instability in model predictions, in contrast to benign examples, which maintain stability despite minor modifications to the inputs. If there's a substantial discrepancy between the transcriptions, the input is labeled as adversarial.
Ma et al.~\cite{ma2021detecting} explore an adversarial sample detection method based on the temporal correlation for audio and video. They believe that the addition of adversarial noise can disrupt the correlation between audio and video. Therefore, a synchronization confidence score is used as a proxy to measure the correlation between audio and video, detecting adversarial samples.
Kwon et al.~\cite{kwon2023audio} introduce a defense method for detecting adversarial examples that does not require a separate module or an additional process. Specifically, they suggest that when optimized noise is embedded into target samples, it leads to a specific pattern in the classification scores of the adversarial example. Experimental results show that this characteristic can be used to identify audio adversarial examples and defend against adversarial attacks.

To detect audio adversarial examples, Guo et al.~\cite{guo2023towards} propose a universal defense algorithm based on audio fingerprint analysis. Specifically, they analyze the similarity between the current query and a set of past queries of a specified length to determine when a series of queries might be susceptible to producing audio adversarial samples.
Park et al.~\cite{park2024toward} introduce an adversarial sample detection algorithm, which inputs both smoothed and original samples into the ASR system and introduces carefully selected noise to logits before decoding. According to their analysis, this carefully selected noise significantly affects the transcription results of adversarial samples but has a minimal impact on clean samples. Therefore, they leverage this characteristic to identify adversarial samples.
Rabhi et al.~\cite{rabhi2024audio} demonstrate that the audio deepfake detection system is vulnerable to adversarial examples. To counter these threats, they introduce a highly generalizable defense mechanism that includes a speech-to-text mechanism. Specifically, if the audio is classified as real by the deepfake detector, the speech-to-text function evaluates whether the audio content matches the expected text. If the audio successfully passes this speech-to-text verification layer, it is then deemed authentic.

\noindent {\bf Optimization }
Esmaeilpour et al.~\cite{esmaeilpour2019robust} introduce a defense algorithm against adversarial attacks. They first demonstrate the robustness of support vector machines (SVMs) when facing several state-of-the-art adversarial attacks. Then, they design a new method based on audio signal preprocessing with Discrete Wavelet Transform (DWT) representations and SVM to protect audio systems from adversarial attacks, which provides a viable balance between accuracy and robustness.
Lu et al.~\cite{lu2023adversarial} enhance the adversarial robustness of audio classifiers through three aspects. First, they demonstrate that $\ell_2$ norm perturbations can generate perturbed examples with specific signal-to-noise ratios. Second, the APGD method is introduced for adversarial training, which enhances the model's robustness against adversarial attacks. Lastly, they leverage data augmentation strategies, such as CutMix, to optimize the model's robustness.
Sun et al.~\cite{sun2024trustnavgpt} leverage MLLMs to build an audio-guided navigation agent, which evaluates the credibility of human instructions based on emotional cues in spoken communication, such as tone and intonation variations. Experimental results indicate that the system possesses remarkable resilience when facing adversarial attacks.

\noindent {\bf Purification }
B. Raj et al.~\cite{olivier2021high} present a more robust adversarial attack defense algorithm than naive noise filtering. Specifically, they utilize a high-pass filter on additive Gaussian noise to smooth the model where it is most vulnerable.
Chen et al.~\cite{chen2022towards} provide valuable insights for enhancing the security of speaker recognition systems through a comprehensive evaluation of various transformation and training methods. They also introduce a new feature compression technique, named FeCo, which compresses and aggregates audio features using clustering methods to mitigate adversarial perturbations, ensuring the security of speaker recognition systems.
Wu et al.~\cite{wu2023defending} introduce an adversarial purification-based defense algorithm based on off-the-shelf diffusion models. This algorithm adds a small amount of noise to the adversarial audio, then runs the reverse sampling step to purify the noisy audio and recover clean audio. It is plug-and-play and can be quickly transferred to other models without the need for retraining.
Du et al.~\cite{du2024adaptive} introduce an adaptive unified defense framework tailored to adversarial attacks. Specifically, they design a unified pre-processing mechanism, which includes RIR convolution, multi-fragment noise padding, and SPL complexity analysis, to disrupt adversarial attacks. Secondly, they leverage multiple automatic speech recognition systems to transcribe pre-processed audio to evaluate similarity and detect adversarial properties. Experimental results show that the framework effectively defends against various adversarial attack strategies.

\begin{table}[htb]
\centering
\begin{tcolorbox}
{\color{blue}{\bf Insights: }The research discussed highlights various adversarial sample detection and purification strategies, such as noise padding and audio fingerprint analysis, which enhance the adaptability and robustness of audio-visual models against different types of attacks. Despite the multitude of defense algorithms, these methods lack sufficient generalizability when facing various types of attacks. Additionally, defending against adversarial attacks requires substantial computational resources, which reduces their practicality.}
\end{tcolorbox}
\end{table}

\vspace{-2mm}
\subsection{Backdoor attack}
\noindent {\bf Sample detecting }Building upon STRIP~\cite{gao2019strip}, a backdoor attack defense algorithm targeting images, Xin et al.~\cite{xin2024speechguard} introduce SpeechGuard to defend against backdoor attacks in the audio domain. Specifically, they optimize the STRIP algorithm by enhancing the perturbation techniques, making it more effective for detecting poisoned audio samples. Furthermore, they utilize time-frequency masking to suppress trigger signals and purify the poisoned samples.
Guo et al.~\cite{guo2023masterkey} introduce a ``sniper"-based backdoor attack defense algorithm that examines the dataset before training to filter out suspicious samples. Specifically, they use the average embedding of the dataset as a ``sniper" and calculate the $\ell_2$ distance between it and other samples to identify potentially malicious samples.

\noindent {\bf Optimization }Zhou et al.~\cite{zhou2025gradient} discover that backdoored neurons exhibit greater gradient values compared to other neurons. Based on this observation, they propose a gradient norm-based fine-tuning algorithm to defend against backdoor attacks. Specifically, they apply gradient norm regularization to fine-tune the victim model, thereby weakening the backdoored neurons and enhancing the model's defense against backdoor attacks.

\noindent {\bf Purification }Zhu et al.~\cite{zhu2022backdoor} introduce a backdoor attack defense algorithm for voice print recognition models based on speech enhancement and weight pruning. Specifically, this algorithm leverages perturbation detection to distinguish between clean and poisoned samples. Subsequently, clean samples are used to fine-tune the model with pruning, and a speech enhancement algorithm is applied to purify the poisoned samples, thereby preventing the activation of the backdoor.
Chen et al.~\cite{chen2023knowledge} present a backdoor attack defense algorithm in federated learning. Specifically, they detect and remove features of the trigger during inference based on knowledge distillation. The knowledge distillation algorithm is used to train a validation model on each IoT device to identify suspicious poisoned samples, and a feature cancellation mechanism is employed to eliminate the trigger features in these suspicious samples, thereby defending against backdoor attacks.
Wu et al.~\cite{wu2024web} design a backdoor attack defense algorithm for federated learning systems, which is based on speaker frequency. This algorithm first filters out infrasound frequencies below 18 Hz and ultrasonic frequencies above 20,000 Hz, aiming to eliminate commonly used backdoor attack triggers. Then, by filtering out frequencies outside the normal human vocal range, they ensure that the integrity and quality of the voice data are preserved, and effectively prevent the activation of backdoors.
Bartolini et al.~\cite{bartolini2024hidden} explore leveraging the voice activity detection model as a defense mechanism against backdoor attacks, aiming to filter out backdoor triggers and prevent the backdoor from being activated. Experimental results indicate that this model is capable of defending against backdoor attacks.

\begin{table}[htb]
\centering
\begin{tcolorbox}
{\color{blue}{\bf Issues to Consider: }Due to the fixed trigger patterns inherent in backdoor attacks, these patterns can serve as distinctive signals for identification and subsequent defense mechanisms. However, the challenge in defending against such attacks extends beyond merely detecting poisoned audio samples. It is equally crucial to ensure that clean audio-visual samples remain unaffected. This dual requirement significantly amplifies the complexity of implementing effective defense strategies against backdoor attacks.}
\end{tcolorbox}
\end{table}

\vspace{-5mm}
\subsection{Other}
Yu et al.~\cite{yu2023antifake} introduce AntiFake, a defense strategy to prevent unauthorized speech synthesis. Concretely, AntiFake disrupts speech synthesis by deviating speaker embeddings used for speaker identity control. Furthermore, to ensure transferability, they also leverage ensemble learning to enhance the generalizability of the AntiFake algorithm.
\vspace{-3mm}
\section{Application}\label{application}
Much like a coin has two sides, various attacks on audio-visual-based multimodal models can serve both as a threat and as an effective tool for evaluation. This dual role facilitates activities such as model copyright auditing and privacy protection, highlighting the complexity and multifaceted nature of security in audio-visual-based multimodal models.

\noindent {\bf Copyright protection} To protect the copyright of speech recognition models, Liao et al.~\cite{liao2024imperceptible} introduce imperceptible backdoor watermarks to authenticate ownership. They utilize Gaussian noise watermarks, extreme frequency Gaussian noise watermarks, and unrelated audio watermarks to embed backdoors into the target model. This enables black-box verification of the intellectual property of the model owners.
Wu et al.~\cite{wu2020audio} propose a steganography method based on adversarial examples, which involves making subtle perturbations to audio in the time domain to evade detection by CNN steganalysis. This method does not rely on existing embedding costs but instead starts with random or simple embeddings, enhancing security performance through iterative cost updates.
Chen et al.~\cite{chen2023model} introduce a model access control scheme based on hidden adversarial samples, focused on the intellectual property protection of automatic speech recognition models. This method utilizes audio adversarial samples by embedding user identity information into the adversarial samples of the audio, serving as proof samples for authentication.

\noindent {\bf Privacy} To protect user privacy, O’Reilly et al.~\cite{o2022voiceblock} propose VoiceBlock, a system that performs adversarial modifications on user audio streams in real-time, de-identifying user speech to safeguard privacy from automatic speaker recognition systems. VoiceBlock employs deep networks and applies time-varying finite impulse response filters to the outgoing audio stream. These modifications prevent automated systems from identifying users based on their voice, while retaining the speech characteristics unchanged for human listeners.

\noindent {\bf Dataset }To better align fake speech datasets with real-life scenarios, Luong et al.~\cite{luong2025llamapartialspoof} hconstruct LlamaPartialSpoof, a 130-hour dataset that contains both fully and partially fake speech. This dataset is created using MLLMs and voice cloning technologies. Its purpose is to provide a more comprehensive assessment of the vulnerabilities in current countermeasure systems. Cai et al.~\cite{cai2024av} construct a dataset named AV-Deepfake1M. This dataset includes three types of manipulations: video manipulations, audio manipulations, and audio-visual manipulations. It addresses the gap in existing deepfake datasets, which do not include these types of partial manipulations.

\section{Summary and Challenges}\label{challenges}
In this section, we summarize the similarities and differences between various types of attacks and defense algorithms, with the aim of gaining a deeper understanding of model security. Additionally, we provide a summary of the challenges and trends to further encourage research into model security.
\subsection{Summary of attacks and defenses}
\noindent {\bf Similarity Analysis} 
Despite the diverse forms of attack algorithms, their goal is to maliciously manipulate the responses of audio-visual-based multimodal models.
Especially, adversarial and jailbreak attacks can influence the outputs of models through multiple iterative queries.
Therefore, different attack methods exhibit elements that are instructive for reciprocal learning.
For example, perturbations in adversarial attacks have the potential to facilitate jailbreak attacks, thereby disrupting the internal defense mechanisms of models and resulting in the output of harmful content.
Moreover, the defense algorithms against adversarial and backdoor attacks can be categorized into sample detection, optimization, and purification, which can also serve as defensive strategies against jailbreak attacks, particularly through purification.

\noindent {\bf Difference Analysis} 
Unlike adversarial or jailbreak attacks, most backdoor attack algorithms require fine-tuning the model to establish an association between the trigger and the target label, thus, backdoor attacks may consume more computational resources in MLLMs.
Furthermore, alterations in the fine-tuned model weights may detrimentally affect the model's generalization capabilities. Consequently, attackers must meticulously balance the effectiveness of backdoor attacks with the preservation of the model's standard performance. In contrast, the primary focus of adversarial attacks is on refining the generation of perturbations to reduce the frequency of queries.

\subsection{Challenges and trends}
\noindent {\bf Attack Algorithms} Despite the proliferation of audio-visual attack algorithms, several challenges remain:
\begin{itemize}
    \item Facing different MLLM architectures, the generalizability of adversarial perturbations significantly impacts the attack success rate. Although previous research has continuously optimized the generalizability of perturbations, the evolving defense capabilities of MLLMs, which leverage security-aligned optimizations, mean that constructing more effective and generalized perturbations remains an ongoing challenge.
    \item For adversarial and jailbreak attacks, particularly in real-time audio streaming systems, attackers are required to make multiple queries to achieve their ultimate objectives. However, in closed-source models, repeated queries require multiple API calls, which can lead to significant expenses. Therefore, exploring efficient algorithms for generating malicious queries or adversarial perturbations is a worthwhile endeavor.
    \item As the parameter count of models increases, backdoor attack algorithms that depend on fine-tuning are becoming impractical. Therefore, exploring innovative backdoor attack algorithms that either do not require fine-tuning or only utilize parameter-efficient fine-tuning, such as poisoning-based instruction or in-context learning, is encouraged.
    For instance, Zhao et al.~\cite{zhao2024universal} propose ICLAttack method based on in-context learning algorithms, avoiding the fine-tuning of large language models. This method can be adapted for use in audio-visual attacks.
    \item In existing backdoor attack algorithms, to establish an alignment between the trigger and the target label, attackers need to implant triggers and modify sample labels. Although this can achieve a feasible attack success rate, samples with modified labels may be detected by defense algorithms, which reduces the stealthiness of the backdoor attack. Therefore, exploring backdoor attack algorithms that do not require modifying sample labels is a trend.
    \item Moreover, exploring the interpretability of attack algorithms is crucial for understanding the deeper working mechanisms of attacks and can also enhance our understanding of the security vulnerabilities in MLLMs. Specifically, interpretable jailbreak attack algorithms can play a crucial role in identifying and addressing the inherent defensive weaknesses of models.
\end{itemize}

\noindent {\bf Defense Algorithms} 
\begin{itemize}
    \item The purpose of exploring attack algorithms is to identify potential security vulnerabilities in audio-visual-based multimodal models, thereby facilitating the development of effective defense algorithms. However, existing defense algorithms are limited to specific types of attacks, which lack sufficient generalization performance, reducing their effectiveness. Therefore, the exploration of more effective defense algorithms remains an ongoing challenge.
    \item In real-time audio streaming systems, which operate under strict time constraints, this challenge becomes particularly pronounced. In MLLM-based systems, the large number of parameters often results in the majority of the time budget being consumed by model inference, leaving only a limited window for rejecting or filtering malicious inputs. This restriction may compromise the effectiveness of defense algorithms. Striking a balance between model efficiency and security thus emerges as a critical challenge for real-time audio streaming systems.
    \item Effective defense algorithms are essential not only for the accurate identification of poisoned audio samples but also for minimizing false positive rates, which are crucial for maintaining the normal performance of models. Particularly in the context of sample detection algorithms, the ability of defense mechanisms to precisely differentiate between poisoned and original audio samples is fundamental to the overall reliability and stability of the model.
    \item Although optimization-based algorithms demonstrate notable efficacy in defending against attacks, they require the fine-tuning of the victim model, which in turn demands substantial computational resources, particularly within the framework of MLLMs. Therefore, reducing the consumption of computational resources during the defense phase and enhancing the efficiency of backdoor mitigation are imperative for evaluating the practicality of defense algorithms.
    \item The absence of defense algorithms against jailbreak attacks targeting audio-visual-based MLLMs presents an urgent challenge. A feasible approach is to leverage defense algorithms from the image or text domains. For example, Wang et al. define the defense against jailbreak attacks as ``backdoor attacks'', where prefixed safety examples with a secret prompt serve as ``triggers''. By fine-tuning, they establish an alignment between the triggers and safety responses, thus preventing the output of harmful content to defend against jailbreak attacks. The aforementioned algorithm can be adapted to defend against jailbreak attacks targeting audio-based tasks. 
    
\end{itemize}

\section{Conclusion}\label{conclusion}
In this paper, we systematically review various attack and defense algorithms in the audio-visual domain. 
Simultaneously, we focus our attention on multimodal large language models (MLLMs), which reveal the potential security vulnerabilities of MLLMs. 
Our research reveals that state-of-the-art audio-visual-based multimodal models are susceptible to adversarial perturbations or malicious queries, resulting in the output of incorrect or harmful content.
In addition, we demonstrate existing defense algorithms against attacks, with sample detection, optimization, and purification.
Finally, we highlight the potential challenges and emerging trends in audio-visual attacks.
We hope that this survey serves as a valuable resource for researchers and practitioners, fostering the security of MLLMs and building a reliable audio-visual community.

	
%
%
%
\bibliographystyle{splncs04}
\bibliography{main}
	
\end{document}